\documentclass[
reprint,        
aps,
prb,
superscriptaddress,
]{revtex4-2}

\usepackage[utf8]{inputenc}
\usepackage{graphicx}
\usepackage{amsmath,amssymb}
\usepackage{booktabs}
\usepackage{float}
\usepackage{xcolor}
\usepackage{hyperref}
\usepackage{caption}

\hypersetup{
	colorlinks=true,
	linkcolor=blue,
	citecolor=blue,
	urlcolor=blue
}

\begin{document}
	
	\title{Semimetallic Superconductivity in Cubic Nd$_3$In:\\
		A First-Principles Insight into Indium-Based Compounds}
	
	\author{Arafat Rahman}
	\affiliation{Department of Physics, University of Dhaka, Dhaka 1000, Bangladesh}
	
	\author{Alamgir Kabir}
	\email{alamgir.kabir@du.ac.bd}
	\affiliation{Department of Physics, University of Dhaka, Dhaka 1000, Bangladesh}
	
	\author{Tareq Mahmud}
	\email{tareqphy1205@gmail.com}
	\affiliation{Department of Physics, University of Dhaka, Dhaka 1000, Bangladesh}
	
	\begin{abstract}
		The quest for materials that simultaneously exhibit superconductivity and nontrivial topology has drawn significant attention in recent years, driven by their potential to host exotic quantum states. Their unique coexistence often leads to rich physics and potential applications in quantum technologies. Here, we predict cubic Nd$_3$In as an exceptional candidate in this class, combining strong-coupling superconductivity with distinctive topological features. Using first-principles calculations, we find that the strong-coupling superconductivity in Nd$_3$In arises primarily due to pronounced Fermi surface nesting, leading to an electron-phonon coupling constant of $\lambda = 1.39$. Our fully anisotropic Migdal--Eliashberg analysis predicts a superconducting transition temperature \( T_c \approx 14\ \mathrm{K} \) at ambient pressure, which is the highest value reported so far among cubic semimetallic superconductors. When subjected to a pressure of 15 GPa, \( T_c \) increases further to 18 K. Beyond superconductivity, Nd$_3$In is found to be a Weyl semimetal, as evidenced by the presence of Fermi arcs and nontrivial $\mathbb{Z}_2$ topological invariants, confirming its topological nature. The combination of strong-coupling superconductivity and nontrivial topological states makes Nd$_3$In a promising candidate for quantum transport and topological quantum computation.
		
	\end{abstract}
	
	\maketitle
	
	\section{Introduction}
Semimetallic superconductors lie at the forefront of condensed matter physics, where the interplay between topological states and electron–phonon interactions gives rise to novel quantum phenomena \cite{yan2020vortex,yuan2019evidence,meng2017erratum,li2018topological,wang2020fractional,veyrat2023berezinskii,kealhofer2020topological,wei2014odd,huang2019proximity,dong2022superconductivity}.
These superconductors not only host Majorana fermions which are crucial for quantum information technologies and topological quantum computation protected from decoherence \cite{ripoll2022quantum,liu20192d,sato2017topological,liang2024polymorphism,krasnok2024superconducting,kitaev2001unpaired,nayak2008non,sau2010generic,freedman2003topological,tang2022high,zhang2023topological} but also give rise to a variety of intriguing quantum phenomena, including topological quantum spin liquids \cite{kasahara2018majorana,wolter2022spin,fu2008superconducting}, the quantum Hall effect \cite{molina2018surface,uchida2017quantum,fu2022quantum,bednik2016anomalous}, quantum transport \cite{lu2017quantum,wang2013three,song2017topological}, and potential applications in high-speed electronics \cite{sufyan2023topological}. Both layered and bulk materials have been found to host semimetallic superconductivity, expanding the landscape for exploring exotic quantum phases \cite{adam2021superconductivity,vocaturo2024electronic,salis2021heat,leng2017type,rosenstein2023superconductivity,chan2017nearly,zhang2020superconductivity}.  Among layered materials, W$_2$N$_3$ stands out with a predicted superconducting transition temperature $T_c$ of 21–28 K, which is one of the highest reported in semimetallic systems \cite{campi2021prediction}. Unlike layered materials, bulk compounds such as AuCu$_3$-type intermetallics are known for their simple cubic symmetry and the versatility of their intriguing properties, such as magnetic ordering \cite{shenoy1970magnetic}, heavy fermion behavior \cite{du2025hot,lu2021temperature}, topological characteristics \cite{yin2022electronic,yang2024realization,cheng2020pressure} and superconductivity \cite{luo2015superconductivity,baǧci2023influence,tu2025superconductivity}. Beyond superconductivity, many AuCu$_3$-type intermetallics also display topological superconductivity. For instance, Sn$_3$Ca ($T_c = 4.2$ K), a superconducting semimetal \cite{siddiquee2022nematic} known for its potential in quantum computing \cite{siddiquee2021fermi}, whereas Sn$_3$Y ($T_c = 7$ K) hosts nontrivial band topology \cite{kawashima2010superconducting,tu2019topological}, and Sn$_3$La ($T_c = 5.9$–6.3 K) displays similar traits under applied pressure \cite{singh2019superconducting}. Additionally, the Tl$_3$Y and Pb$_3$Y compounds exhibit Rashba-like surface states and phonon-mediated superconductivity, with transition temperatures of $T_c = 4.37$ K and 2.17 K, respectively \cite{tu2019topological}. These low $T_c$ superconductors have become integral to technologies such as fusion reactors, particle accelerators, magnetic resonance imaging (MRI), and superconducting radio frequency (SRF) cavities \cite{banno2023low}. For example, Nb$_3$Sn, with a transition temperature of $T_c = 18.3$ K \cite{xu2017review}, is widely utilized in the aforementioned technologies \cite{posen2022nb3sn,buta2021very,miyoshi66,mentink2014experimental}.
Moreover, recent experiments have been able to detect the presence of topologically protected surface states in superconductors, including the observation of chiral surface states \cite{yao2024observation,deng2024chiral,ming2023evidence} and Fermi arcs in semimetallic systems \cite{kuibarov2024evidence,yang2023coexistence}. These discoveries have significantly advanced the field, expanding the frontier of both fundamental physics and practical applications. 
\\
Despite Indium (In) being a well-known elemental superconductor, research on In-based superconducting compounds has remained limited. Most available studies were conducted several decades ago and include compounds such as La$_3$In ($T_c = 9.5$ K) \cite{zhao1995synthesis}, In$_3$Ce ($T_c = 0.2$ K) \cite{fukazawa2003theory}, InHg ($T_c = 3.1$–4.5 K) \cite{merriam1963superconductivity}, InSn ($T_c = 3.4$–6.6 K) \cite{merriam1963superconductivity2}, InTl ($T_c = 2.5$–3.8 K) \cite{merriam1967superconductivity3}, and InSb ($T_c = 1.88$ K) \cite{stromberg1964superconductivity}. This lack of recent investigation reveals a significant gap in the study of In-based superconductors, especially in the context of modern materials discovery. Furthermore it has been observed that rare earth substitutions particularly involving Neodymium (Nd) have demonstrated notable success in enhancing $T_c$ across various systems \cite{jassim2019enhancement,biju2007structural,cheng2009high}. This suggests that the unique combination of In and Nd in Nd$_3$In could provide a promising route for improving superconducting properties. Moreover, X-ray diffraction analysis indicates that Nd$_3$In crystallizes in a cubic structure \cite{moriarty1966x}, aligning it structurally with the class of previously discussed AuCu$_3$-type cubic superconductors.  The presence of heavy atoms such as Nd and In in Nd$_3$In is expected to produce strong spin–orbit coupling effects that significantly influence its electronic structure and give rise to topological phases, while also enabling spintronic applications \cite{zhang2025nodal}. Exploring the superconducting and topological aspects of Nd$_3$In not only revives interest in In-based superconductors but also contributes to the broader search for high-performance semimetallic superconductors within the AuCu$_3$-type family.
In this study, we present a comprehensive theoretical investigation of the electronic, phononic, superconducting, and topological properties of cubic Nd$_3$In. The study begins with the analysis of the electronic structure including spin–orbit coupling in Section A, followed by phonon calculations to investigate lattice dynamics and possible soft modes in Section B. In Section C, we evaluate the Fermi surface nesting and electron–phonon coupling to uncover the microscopic origin of superconductivity in Nd$_3$In. Finally, Section D is devoted to exploring its topological characteristics.
\section{ Methodology and Crystal structure}
\begin{figure}[h!]
    \centering
    \includegraphics[width=1\linewidth]{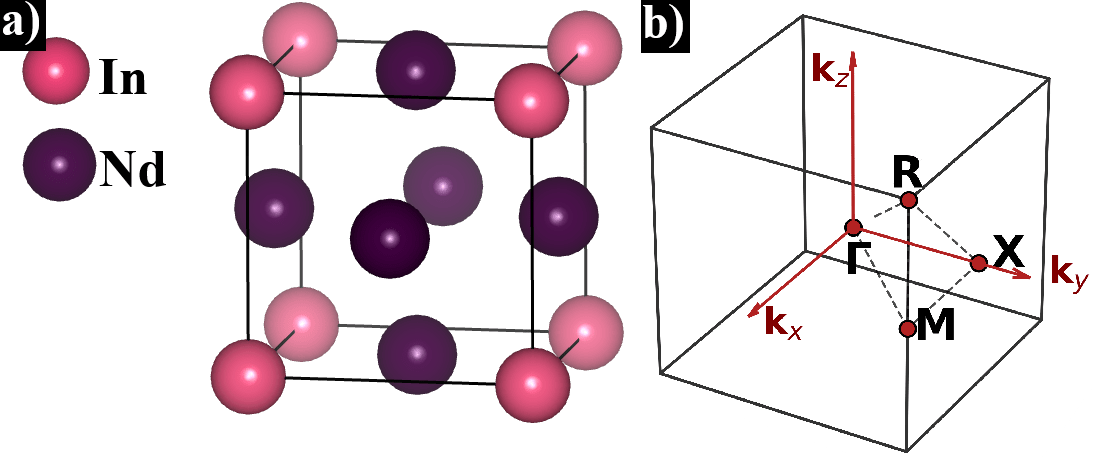}
    \caption{(a) Crystal structure and (b) Brillouin zone of Nd$_3$In.}
    \label{fig:structure}
\end{figure}

The crystal structure of Nd$_3$In belongs to the primitive cubic space group Pm$\bar{3}$m (No. $221$), with a calculated lattice parameter of 4.977 \AA (4.982 $\text{\AA}$ with SOC), which are in close agreement with the experimentally reported value \cite{moriarty1966x,yatsenko1983phase}. In this structure, the In atom occupies the high-symmetry 1a Wyckoff position at the origin (0.0, 0.0, 0.0), while the Nd atoms are located at the 3c Wyckoff positions with fractional coordinates (0.0, 0.5, 0.5), (0.5, 0.0, 0.5), and (0.5, 0.5, 0.0). The optimized crystal structure of Nd$_3$In and its corresponding Brillouin zone are illustrated in Figure \ref{fig:structure}(a) and Figure \ref{fig:structure}(b), respectively.
\vspace{4mm}
\begin{figure*}
	\centering
	\includegraphics[width=1\linewidth]{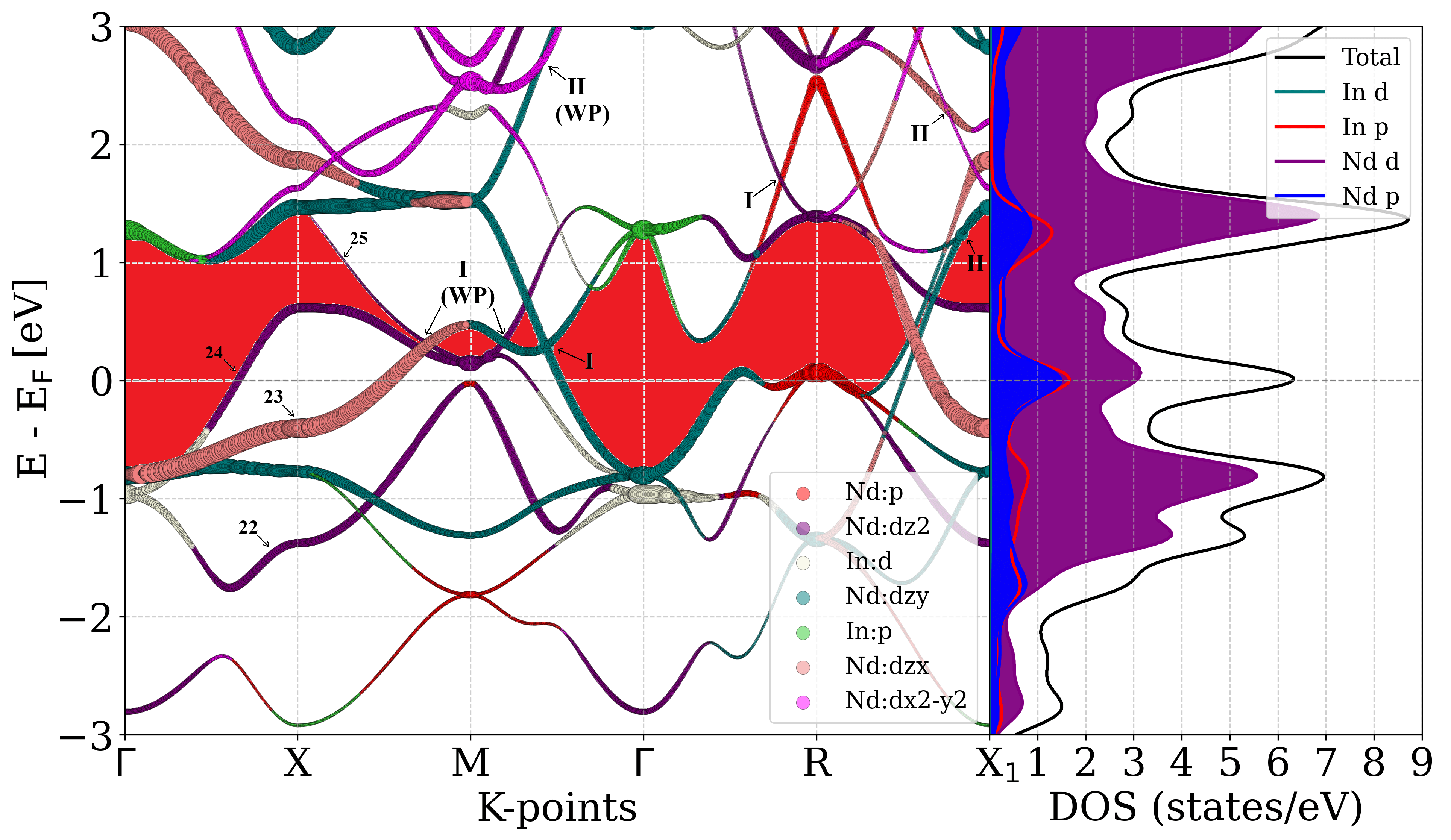}
	\captionsetup{justification=raggedright,singlelinecheck=false}
	\caption{Orbital projected electronic band structure and corresponding density of states (DOS) for Nd$_3$In. The Fermi energy ($E_F$) is set to zero and indicated by the dashed horizontal line.  The size of the colored scatter points is proportional to the contribution from the selected atomic orbitals. Several Weyl points are observed in the band structure, among which a few representative points are highlighted. The red-shaded region in the left panel marks the locations of type-I Weyl points, which occur between bands 24 and 25.}
	\label{fig:banddos}
\end{figure*}
\begin{figure}
	\centering
	\includegraphics[width=1\linewidth]{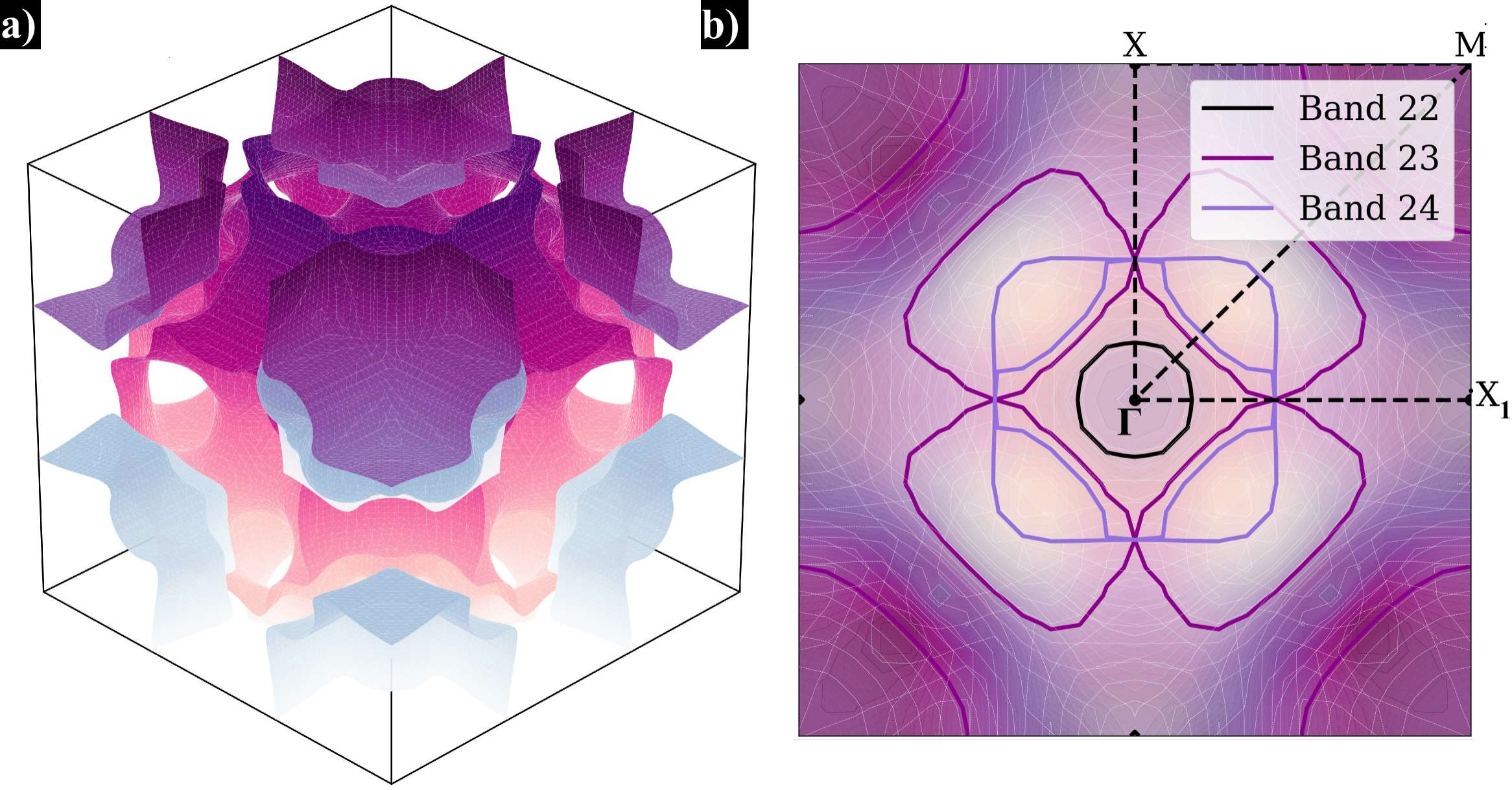}
	\captionsetup{justification=raggedright,singlelinecheck=false}
	\caption{(a)  Fermi surface in reciprocal space. (b) Contours of the Fermi surface on the $k_z = 0$ plane, with dashed lines indicating the high symmetry points.}
	\label{fig:fermi}
\end{figure}
\begin{figure*}
	\centering
	\includegraphics[width=1\linewidth]{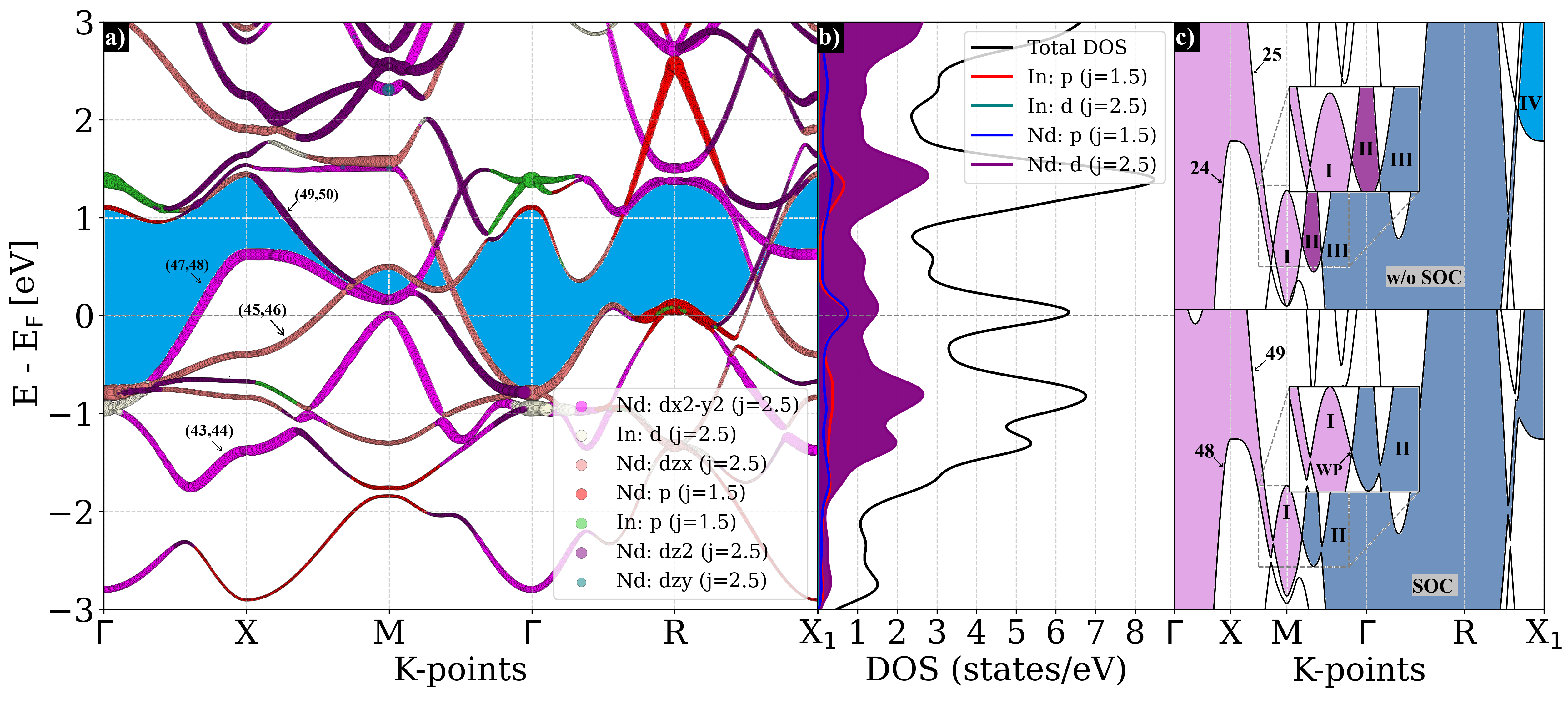}
	\captionsetup{justification=raggedright,singlelinecheck=false}
	\caption{(a) SOC-induced orbital-projected electronic band structure of Nd$_3$In and (b) the corresponding density of states (DOS). The Fermi energy ($E_F$) is set to zero and indicated by the dashed horizontal line. The size of the colored scatter points in panel (a) reflects the relative contribution of the selected atomic orbitals. (c) Effect of SOC on the band structure. Without SOC (upper inset), three band-touching points divide the spectrum into four regions (I–IV). Inclusion of SOC (lower inset) lifts two of these nodal points, leaving a single Weyl point and partially gapped regions.}
	\label{fig:bandosrel}
\end{figure*}
\\First-principles calculations based on density functional theory (DFT) are performed using the plane-wave pseudopotential code \texttt{Quantum ESPRESSO} \cite{giannozzi2009quantum}. The exchange-correlation potential is treated within the generalized gradient approximation (GGA) using the Perdew-Burke-Ernzerhof (PBE) functional. Scalar-relativistic ultrasoft pseudopotentials are used for both Nd and In atoms. For Nd, the 4f electrons are strongly localized and therefore excluded from the valence states. They are treated as frozen core states following the standard open-core approach, since conventional DFT does not provide a reliable description of localized f electrons\cite{gu2020substantial,liu2024segregation,ferrari2024density}. The valence electron configurations considered are $4d$, $5s$, and $5p$ for In, and $5d$, $6s$, and $6p$ for Nd. A kinetic energy cutoff of 75~Ry is used for the plane-wave basis set, with a charge density cutoff of 350~Ry. The Brillouin zone is sampled using a $\Gamma$-centered $24 \times 24 \times 24$ \textbf{k}-mesh. Gaussian smearing with a width of 0.0125~Ry is applied for electronic occupations. Structural relaxations are carried out until the total energy and atomic forces converge below $1.0 \times 10^{-10}$~Ry and $1.0 \times 10^{-6}$~Ry/Bohr, respectively.
Phonon dispersions, electron-phonon interactions, and superconducting parameters are evaluated using density functional perturbation theory (DFPT). For accurate estimation of superconducting properties, the \texttt{EPW} package is employed, which uses Wannier interpolation of electron-phonon matrix elements to efficiently calculate key quantities on dense grids \cite{ponce2016epw}. It should be noted that EPW cannot be applied within DFT+U calculations, which are often required for localized f systems. Furthermore, the use of hybrid functionals would improve the electronic description but is computationally expensive and beyond the scope of this study. The superconducting transition temperature $T_c$ is estimated using the Allen-Dynes modification of the McMillan formula \cite{allen1975transition}:
where $\omega_{\log}$ denotes the logarithmic average of phonon frequencies, and $\mu^*$ is the retarded Coulomb pseudopotential. The correction factors $f_1$ and $f_2$ are expressed as
with $\omega_2$ being the mean-square phonon frequency defined by
 The total electron-phonon coupling (EPC) constant $\lambda$ is given by
where $\alpha^2F(\omega)$ is the Eliashberg spectral function, defined as
Here,  are the electron-phonon matrix elements, and are the phonon frequencies.
To further investigate the electronic structure, the Fermi surface nesting function $\zeta(\mathbf{q})$ is computed as
where $\Omega_{\text{BZ}}$ is the Brillouin zone volume and $E_F$ is the Fermi energy. 
Topological properties, including surface states and invariants, are computed using \texttt{WANNIER90} \cite{mostofi2008wannier90} and \texttt{WANNIERTOOLS} \cite{wu2018wanniertools}. Maximally localized Wannier functions are constructed with spin-orbit coupling (SOC), and surface Green’s function methods are used to resolve the surface band structure. Topological invariants are evaluated from the Wannier-based tight-binding model.
\section{Results and Discussion}
\subsection{Electronic  properties}
The orbital projected bandstructure of Nd$_3$In (without SOC), alongwith its DOS is shown in Figure \ref{fig:banddos}. 
\begin{figure}
	\centering
	\includegraphics[width=1\linewidth]{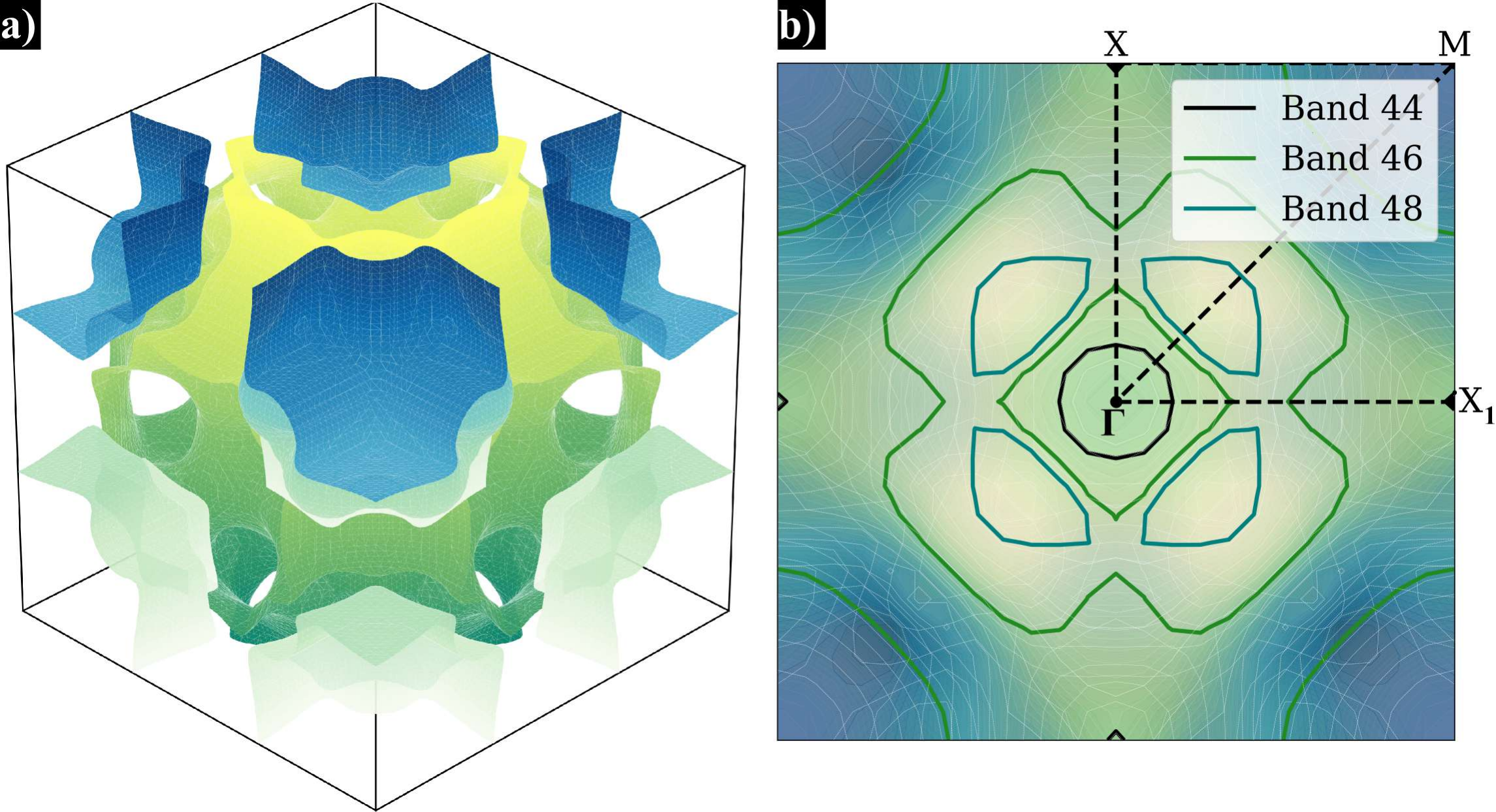}
	\captionsetup{justification=raggedright,singlelinecheck=false}
	\caption{(a) Fermi surface including spin--orbit coupling (SOC) displayed in reciprocal space. (b) Cross-sectional contours of the Fermi surface on the $k_z = 0$ plane, where dashed lines denote high-symmetry points.}
	\label{fig:fermirel}
\end{figure}
As shown in the right panel of Figure \ref{fig:banddos}, the Nd $d$ states dominate the electronic structure across the entire energy range. Notably, the primary contribution at the Fermi level comes from the Nd $d_{z^2}$ orbitals, followed by the $d_{zx}$, $d_{zy}$ and Nd $p$ orbitals, as these are the most saturated near the Fermi energy. The left panel of Figure \ref{fig:banddos} illustrates bands 22, 23, and 24 intersect the Fermi level, resulting in the formation of three distinct Fermi surfaces, all displayed together in a single plot in Figure \ref{fig:fermi}(a).
The Fermi surface arising from band 22 forms a hole pocket around the $\Gamma$ point, primarily composed of Nd $p$ orbital character and outlined by the black contour as illustrated in Figure \ref{fig:fermi}(b). Band 23 gives rise to hole-like pockets near the M point and four electron-like pockets along the M-\(\Gamma\) direction, as indicated by the purple contours in Figure \ref{fig:fermi}(b). These hole pockets near the M point can be explicitly seen in Figure \ref{fig:fermi}(a). These features predominantly originate from Nd \(d_{zx}\) and \(d_{z^2}\) orbitals. Lastly, band 24 forms a nearly rectangular electron-like pocket, indicated by the medium purple contours in Figure \ref{fig:fermi}(b), primarily originating from the Nd $d_{z^2}$ orbital. Within this rectangular pocket, four smaller hole pockets are embedded, which stem from the Nd $d_z{^2}$, $d_{zy}$ and $p$ orbitals. A particularly striking characteristic of the Fermi surface is the presence of several nearly parallel segments, indicating strong Fermi surface nesting, which leads to strong electron–phonon interaction \cite{yang2024phonon,chen2021emergence,chen2022strong}. The nesting vectors can be identified along the $\Gamma$–X, $\Gamma$–M, and $\Gamma$–X$_1$ directions, connecting  flat regions of the Fermi surface. \\
Upon the inclusion of spin–orbit coupling (SOC), six bands (indexed 43 to 48) are observed to cross the Fermi level, with three of them being pairwise degenerate, as illustrated in Figure \ref{fig:bandosrel}(a). The corresponding density of states at the Fermi level, $N(E_F)$, shows a slight decrease from approximately 6.192 eV (without SOC) to 6.118 eV, as presented in Figure \ref{fig:bandosrel}(b), indicating that SOC has a minimal effect on the overall electronic density of states in Nd$_3$In.
 However, SOC significantly affects the band structure of Nd$_3$In by lifting degeneracies and opening finite energy gaps as shown in Figure \ref{fig:bandosrel}(c). The bulk band structure without SOC hosts several Weyl points of both type-I and type-II, a few of which are marked in Figure \ref{fig:banddos}. Type-I Weyl points correspond to band crossings that produce point-like Fermi surfaces, whereas type-II Weyl points occur at the intersection between electron and hole pockets, giving rise to significantly tilted Weyl cones \cite{zhang2025nodal, wang2018type, soluyanov2015type, wang2016observation}.
 When SOC is included, a number of these Weyl crossings are lifted, modifying the bulk electronic spectrum. This effect can be clearly seen in Fig.~\ref{fig:bandosrel}(c). In the case without SOC, three band-touching points appear within the plotted energy range, dividing the spectrum into four distinct regions labeled I–IV. These touching points prevent the opening of a global gap and give rise to apparent partial-gap features. Upon inclusion of SOC, two of these Weyl points are annihilated due to SOC-induced splitting, leaving only a single Weyl point between regions I and II. As a result, the spectrum reorganizes into two main regions, and although partial gaps are present, no full insulating gap opens. The persistence of this Weyl point ensures that Nd$_3$In retains its semimetallic character under SOC rather than becoming a topological insulator. This behavior is consistent with previous reports on SOC in topological semimetals, where SOC often eliminates some Weyl nodes but leaves others intact, thereby stabilizing the Weyl semimetal phase \cite{liu2022direct}. Additionally, the SOC band structure exhibits double degeneracy throughout the Brillouin zone, indicative of Kramers degeneracy \cite{rosch1983time}. This degeneracy indicates the preservation of time-reversal symmetry (TRS) in the system and suggests the possibility of nontrivial topological properties \cite{karn2023band}. These observations indicate the presence of nontrivial topological states in Nd$_3$In, the detailed analysis of which is presented in Section D.\\
 \begin{figure*}
 \centering
 \includegraphics[width=1\linewidth]{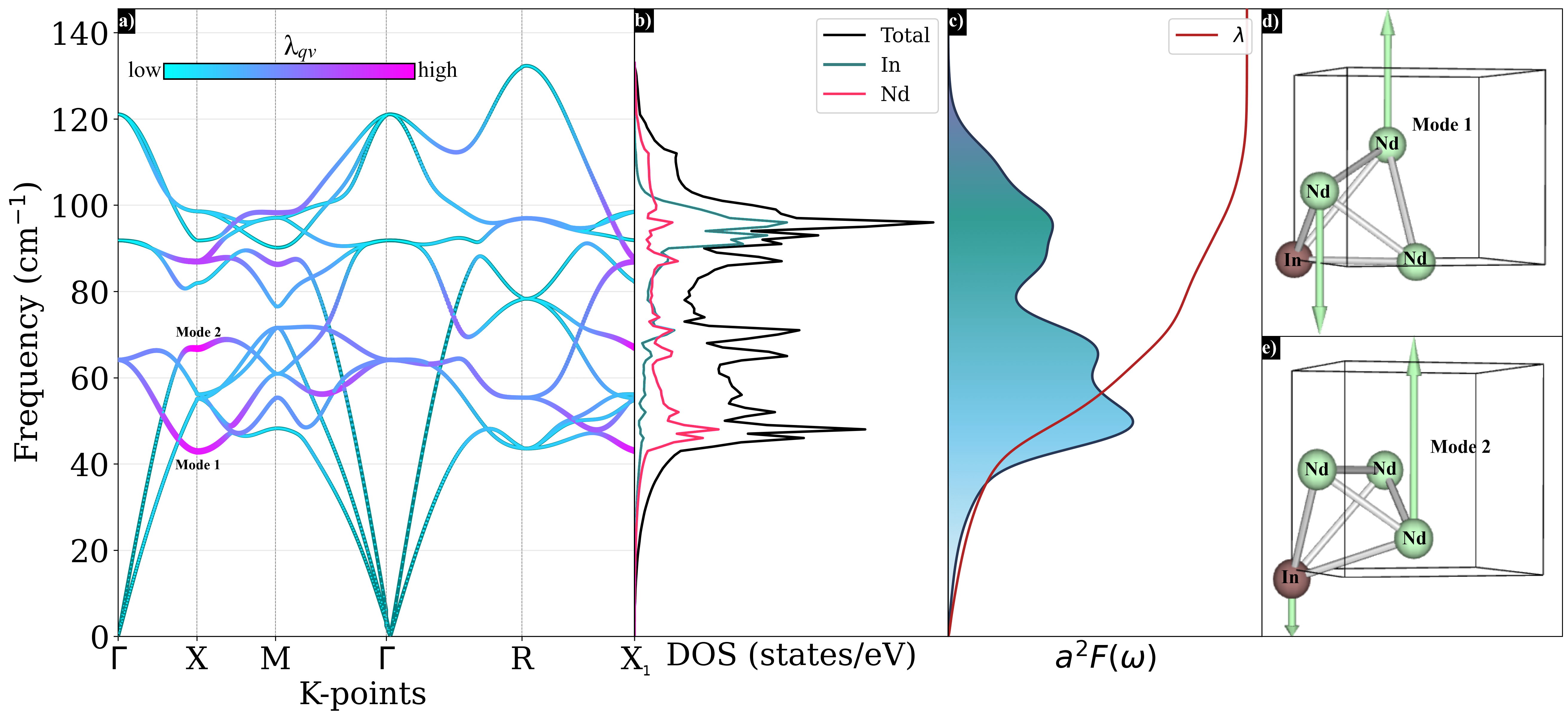}
 \caption{(a) Phonon dispersion relations weighted by the mode-resolved electron–phonon coupling strength $\lambda_{q\nu}$, where magenta shading indicates stronger coupling and cyan shading denotes weaker coupling. (b) Atom-projected phonon density of states (PHDOS). (c) Eliashberg spectral function $\alpha^2F(\omega)$ and cumulative electron–phonon coupling constant $\lambda(\omega)$ as functions of phonon frequency. (d–e) Visualization of the two dominant phonon modes contributing most strongly to the EPC.}
 \label{fig:phona2f}
 \end{figure*}Furthermore, SOC also alters the orbital character and the topology of the Fermi surfaces. The dominant states now arise from the Nd \(d_{x^2 - y^2}\), \(d_{zx}\), and \(d_{z^2}\) orbitals, with a noticeable increase in the contribution from In \(p\) orbitals, which previously played a minimal role in the non-SOC case. The Fermi surfaces corresponding to the three non-degenerate bands (bands 44, 46, and 48) are shown in  \ref{fig:fermirel}(a), while their degenerate partners (bands 43, 45, and 47) exhibit identical features and are therefore not shown separately. Without SOC, the Fermi surface features a prominent rectangular electron-like contour, primarily originating from band 24. After incorporating SOC, this contour disappears and is replaced by lobed teal-colored contours shown in Figure \ref{fig:fermirel}(b). This change arises because bands 46 and 48, which correspond to bands 23 and 24 in the absence of SOC, become non-degenerate near the Fermi level along the \(\Gamma\)–R and R–M directions. These bands now cross the Fermi level at different \(k\)-points, altering the Fermi surface topology.
\begin{figure}
 \centering
 \includegraphics[width=0.85\linewidth]{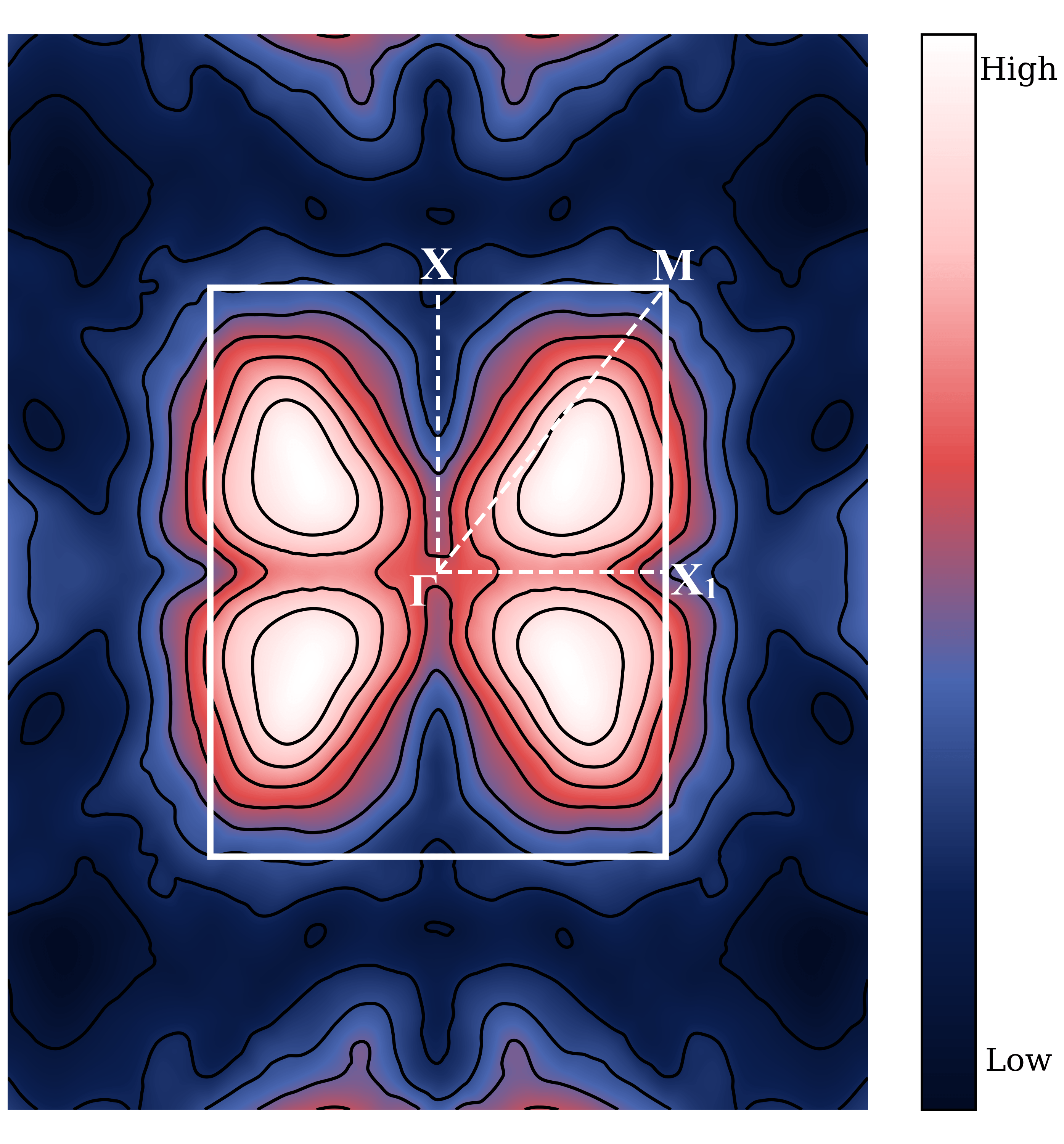}
 \captionsetup{justification=raggedright,singlelinecheck=false}
 \caption{The Fermi surface nesting function $\zeta(\Vec{Q})$} in the $k_z=0$ plane of Nd$_3$In. SOC is not considered.
 \label{fig:nesting}
 \end{figure}
 \begin{figure}
 \centering
 \includegraphics[width=1\linewidth]{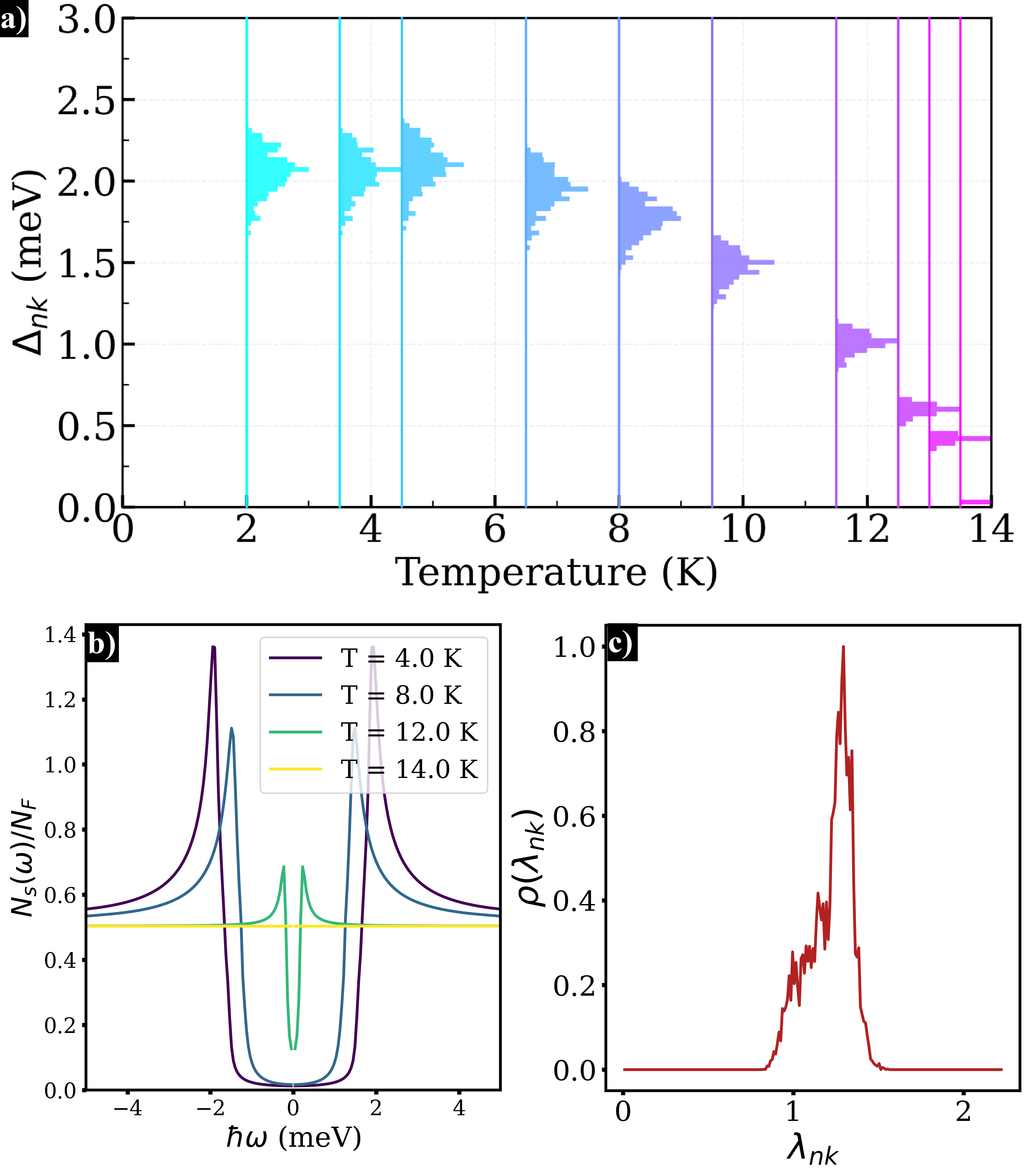}
 \caption{
 	(a) Energy distribution of the anisotropic superconducting gap as a function of temperature. 
 	(b) The quasiparticle density of states (DOS) in the superconducting state compared to the normal-state DOS as a function of energy. The superconducting DOS has been rescaled to match the normal-state DOS, allowing easy comparison with experiments. 
 	(c) The distribution of anisotropic electron-phonon coupling (EPC) strength \(\lambda_\mathbf{k}\), showing a wide variation due to strong anisotropy.
 }
 \label{fig:gap}
 \end{figure} 
\subsection{Phonon dispersion and electron-phonon interaction}
The phonon dispersion and projected phonon density of states (PHDOS) for cubic Nd$_3$In are shown in Figure \ref{fig:phona2f}(a)-(b). The absence of imaginary frequencies throughout the Brillouin zone confirms the dynamical stability of the structure. The phonon spectrum of Nd$_3$In can be divided into three distinct frequency regions. This distribution reflects the expected behavior that heavier atoms like Nd vibrate at lower frequencies than lighter atoms such as In. A closer inspection of the phonon dispersion reveals the presence of soft phonon modes, or Kohn anomalies, along specific high-symmetry directions particularly along the $\Gamma$--X--M and R--X paths. Generally, these soft phonon modes are indicative of strong electron-phonon coupling \cite{jin2020first,jamwal2024enhancement,patel2024electron,sun2022electron}. All three acoustic branches as shown in Figure \ref{fig:phona2f}(a), exhibit noticeable softening along these directions, with the sixth vibrational mode showing pronounced softening along R--X$_1$. We have also examined the pressure dependence of this phonon softening, as shown in Figure S1 of the Supplementary Materials \cite{SM}. The phonon dispersion weighted by mode-resolved electron-phonon coupling strength and the corresponding spectral functions were computed at 10 GPa and 15 GPa pressure. The softening intensifies under applied pressure, suggesting a possible enhancement in the superconducting critical temperature. Figure \ref{fig:phona2f}(c) shows the Eliashberg spectral function $\alpha^2F(\omega)$ and the cumulative EPC function $\lambda(\omega)$.
\\
To understand the microscopic origin of the strong electron--phonon coupling, we analyzed the phonon eigenvectors of the most strongly coupled modes. Two representative modes with the highest EPC values ($\lambda_{q\nu}$) are highlighted in Figure \ref{fig:phona2f}(a). The phonon mode at the X point (Mode 1, $\approx 45$~cm$^{-1}$) is dominated by the vibrations of two Nd atoms oscillating in opposite directions, while the remaining atoms remain stationary as illustrated in Figure \ref{fig:phona2f}(d). The next strongly coupled mode (Mode 2, $\approx 65$~cm$^{-1}$), also located at the X point, involves one Nd atom and In atom vibrating out of phase with unequal amplitudes, while the other atoms remain immobile as shown in Figure \ref{fig:phona2f}(e). This specific relative motion of the atoms strongly perturbs the electronic states near the Fermi energy, accounting for its significant role in the EPC. \\
 To investigate the origin of the Kohn anomalies at ambient pressure, we computed the Fermi surface nesting function $\zeta(\mathbf{Q})$ using the EPW code on a dense $k$-grid of $64 \times 64 \times 64$ and a $q$-grid of $32 \times 32 \times 32$, as shown in Figure \ref{fig:nesting}. The nesting function exhibits pronounced peaks along high-symmetry directions where phonon softening is observed, indicating strong Fermi surface nesting. At the $\Gamma$ point, the prominent peak corresponds to the entire Fermi surface nesting onto itself, which lacks physical significance. Figure \ref{fig:nesting} shows that $\zeta(\mathbf{Q})$ exhibits notably high values along the $\Gamma \rightarrow$ X, $\Gamma \rightarrow$ X$_1$ directions and as well as between the $\Gamma \rightarrow$ M path. As previously discussed, this indicates that parallel segments of the Fermi surface are connected by specific wave vectors $\mathbf{Q}$, in accordance with the features observed in the calculated Fermi surfaces shown in Figure \ref{fig:fermi}(b). Such nesting enhances the electron-phonon interaction, leading to strong-coupling and giving rise to Kohn anomalies \cite{liu2022two,chen2021emergence,kohn1959image,girotto2023dynamical,wang2021high}.
 We then performed a quantitative evaluation of the electron–phonon coupling (EPC) strength. Initially, density-functional perturbation theory (DFPT) calculations on a coarse $4 \times 4 \times 4$ $\mathbf{q}$-mesh yielded a preliminary EPC constant of $\lambda = 0.982$. To obtain a more accurate value, we employed the EPW code, which utilizes maximally localized Wannier functions (MLWFs) to interpolate the electron–phonon matrix elements over dense Brillouin zone grids, specifically a $48 \times 48 \times 48$ $\mathbf{k}$-mesh and a $24 \times 24 \times 24$ $\mathbf{q}$-mesh. To construct the MLWFs, we used initial projections on the Nd $d$ and In $p$ orbitals. The accuracy of the Wannierization was verified by comparing the Wannier interpolated band structure with the original DFT results, as shown in Figure S2 (see Supplementary Materials \cite{SM}).
From the EPW calculations, we obtained a refined electron–phonon coupling (EPC) constant of $\lambda = 1.394$, indicating strong coupling and suggesting that Nd$_3$In is a promising candidate for phonon-mediated superconductivity. The dominant contributions to the total EPC arise from soft phonon modes, as illustrated in Figure \ref{fig:phona2f} (a), where the mode-resolved coupling strengths $\lambda_{q\nu}$ are projected onto the phonon dispersion. A pronounced enhancement of $\lambda_{q\nu}$ is observed along the $\Gamma$--X--M and R--X$_1$ directions, which coincide with regions of phonon softening.
 Analysis of $\lambda(\omega)$ reveals that phonons below 65 cm$^{-1}$ contribute approximately 62.1\% of the total EPC, while intermediate-frequency modes between 65 and 90~cm$^{-1}$ account for 20.8\%. The remaining contribution comes from higher-frequency modes.
\subsection{Superconductivity}
To investigate the superconducting behavior of cubic Nd$_3$In, we begin by examining the nature of its superconducting gap and the strength of electron-phonon interactions. Our analysis reveals that Nd$_3$In exhibits a single-gap superconducting state across the Fermi surface, as shown in Figure \ref{fig:gap}(a). This one gap nature is further supported by a single, sharp peak in the superconducting quasiparticle density of states (QDOS) at different temperatures, illustrated in Figure \ref{fig:gap}(b). Tunneling spectroscopy experiments can directly probe this superconducting QDOS \cite{gu2020single}. Typically, an $s$-wave superconductor shows a U-shaped QDOS, while a $d$-wave superconductor exhibits a V-shaped profile \cite{li2024two}. Our calculated QDOS clearly displays a U-shaped signature, reflecting conventional $s$-wave pairing symmetry in Nd$_3$In.
The electron-phonon coupling (EPC) coefficient in Nd$_3$In exceeds 1.3, indicating a strong-coupling regime. In such cases, the full Allen–Dynes formula is typically employed to estimate the superconducting transition temperature \cite{liu2022two}. Using the widely accepted Coulomb pseudopotential value of $\mu^\star = 0.1$ \cite{pramanick2025pressure,chen2025first,chen2024ambient}, this approach yields a predicted $T_c$ of 9.498 K. However, the Allen-Dynes formula does not account for the momentum dependence of the electron–phonon interaction across the Fermi surface and therefore fails to accurately describe materials exhibiting significant anisotropy \cite{zhao2018multigap}.
To better capture the anisotropic nature of EPC in Nd$_3$In, we  analyze the momentum-dependent EPC strength, defined as .  As shown in Figure \ref{fig:gap}(c), the $\lambda_\mathbf{k}$ values span a wide range from 0.85 to 1.57, clearly indicating strong anisotropy and underscoring the limitations of isotropic approximation in this system. To obtain a more reliable estimate of the superconducting critical temperature $T_c$, we employ fully anisotropic Migdal–Eliashberg calculations, which explicitly incorporate momentum-dependent electron–phonon interactions across the Fermi surface \cite{chen2021emergence,choi2002first,margine2014two}.\\
The temperature evolution of the superconducting energy gap $\Delta_{nk}$ is presented in Figure \ref{fig:gap} (a). At 2 K, the gap ranges from 1.75 to 2.30 meV, reflecting its anisotropic nature. As temperature increases, the gap gradually closes and disappears around 14 K, which we identify as the true superconducting transition temperature. This value shows a 32.16\% enhancement over the isotropic Allen–Dynes estimate, highlighting the significance of momentum-dependent electron-phonon coupling considered in the fully anisotropic Migdal–Eliashberg formalism, as previously mentioned. Notably, Nd$_3$In exhibits the highest superconducting transition temperature among the AuCu$_3$-type semimetallic superconductors. Moreover, $T_c$ further increases under pressure, reaching 16.5~K at 10~GPa and 18~K at 15~GPa, as shown in Figure S3 and S4 (in Supplementary Materials \cite{SM}). Thus, Nd$_3$In can be classified as a low-$T_c$ superconductor and, as previously discussed, holds strong potential for applications in technologies such as MRI, SRF cavities, fusion reactors, and particle accelerators.
 Furthermore, the ratio $2\Delta_0/k_B T_c = 3.93$ exceeds the BCS weak-coupling limit of 3.52, underscoring the strong-coupling character of superconductivity in Nd$_3$In, which primarily originates from the pronounced Fermi surface nesting. These findings clearly demonstrate that the isotropic McMillan–Allen–Dynes approach underestimates $T_c$ in materials with pronounced EPC anisotropy, while the anisotropic Migdal–Eliashberg theory provides a more robust and detailed understanding of the superconducting mechanism.\begin{figure}
	\centering
	\includegraphics[width=1\linewidth]{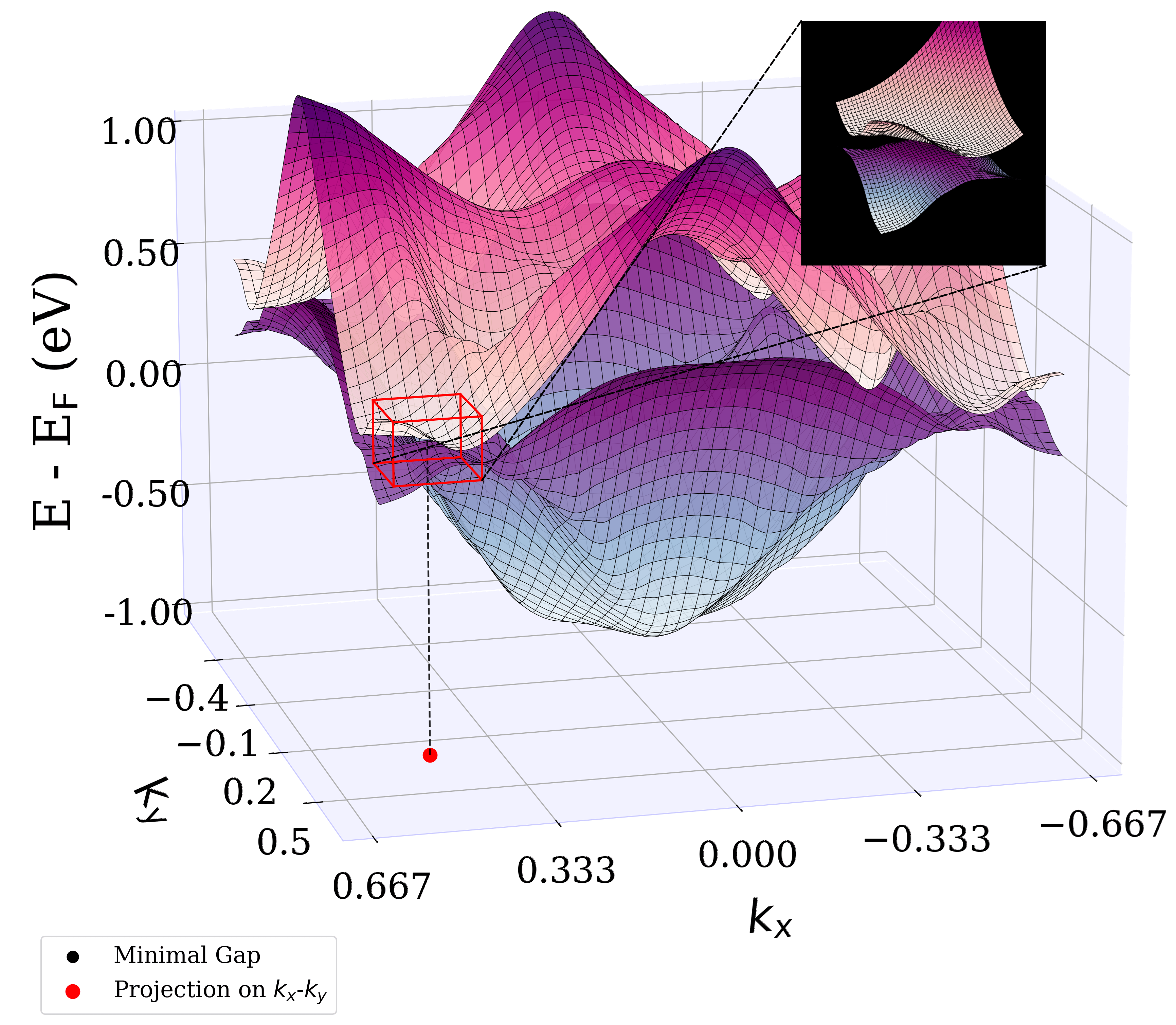}
	\captionsetup{justification=raggedright,singlelinecheck=false}
	\caption{3D plot of the electronic band structure (with SOC) showing the Weyl point located between the topmost valence and lowest conduction bands. The projection of the Weyl point onto the $k_x$–$k_y$ plane is marked with a red dot.}
	\label{fig:bulkek}
\end{figure}
\begin{figure}
	\centering
	\includegraphics[width=1\linewidth]{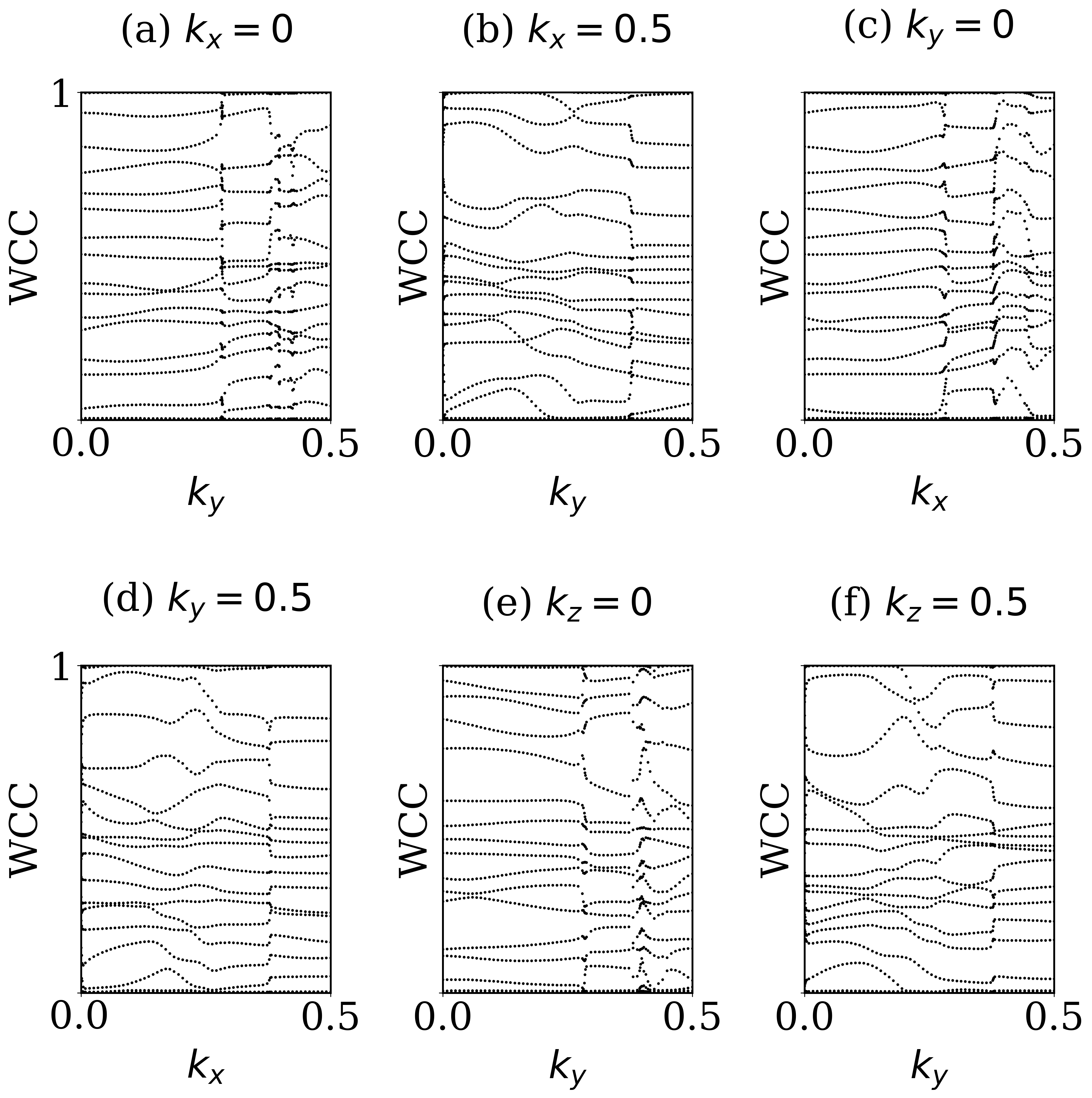}
	\captionsetup{justification=raggedright,singlelinecheck=false}
	\caption{Evolution of Wannier charge centers (WCCs) on different time-reversal invariant momentum (TRIM) planes.}
	\label{fig:WCC}
\end{figure} 
\begin{figure*}
	\centering
	\includegraphics[width=1\linewidth]{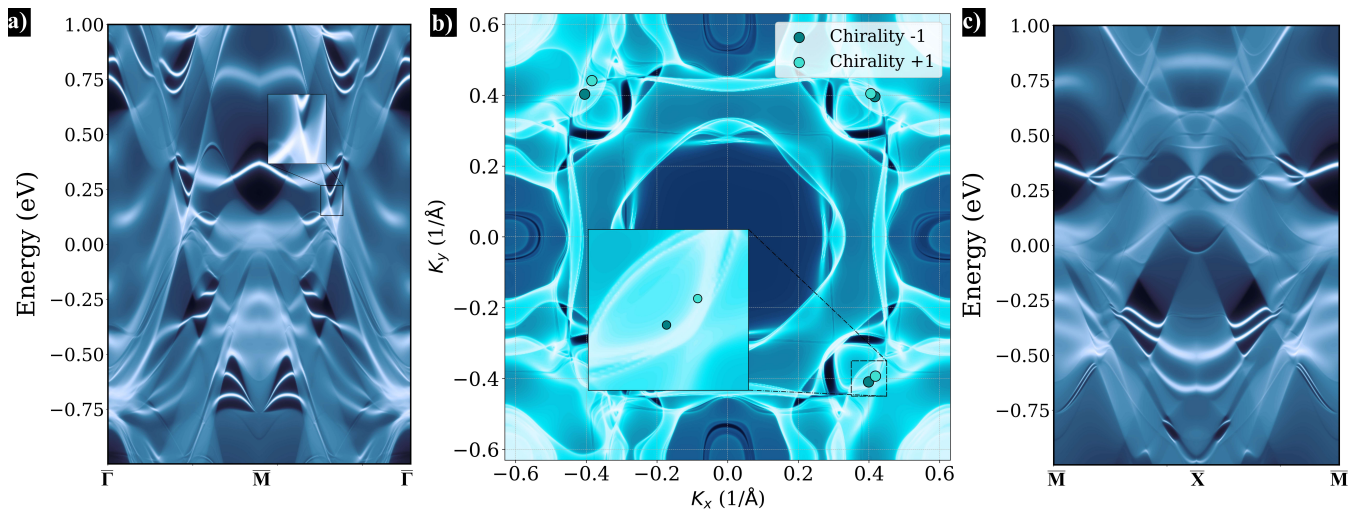}
	\captionsetup{justification=raggedright,singlelinecheck=false}
	\caption{Surface states obtained using the iterative Green's function method based on the Wannierized Hamiltonian. The intensity scale progresses from black, representing low magnitudes, to light bluish-white, indicating high magnitudes.
		(a) Simulated angular-resolved photoemission spectroscopy (ARPES) spectrum along the $\overline{\Gamma}$–$\overline{\text{M}}$–$\overline{\Gamma}$ path near the Fermi level.
		(b) Presence of a Fermi arc verifying the semimetallic behavior. The magnified view highlights the arc connecting Weyl points of opposite chirality.
		(c) Topological surface states (TSSs) along the $\overline{\text{M}}$–$\overline{\text{X}}$–$\overline{\text{M}}$ path.}
	\label{fig:ss}
\end{figure*} \subsection{Semimetallic topology}
As previously discussed, the SOC-induced electronic band structure of Nd$_3$In suggests the presence of nontrivial topological behavior, likely arising from time-reversal symmetry (TRS). As shown in Figure~\ref{fig:bulkek}, the three-dimensional band structure reveals a linear crossing near the $\Gamma$–M path within the energy range of 0.10–0.25 eV, indicating the existence of a type-I Weyl point formed between the lowest conduction band and the highest valence band.  Our primary focus remains on this particular crossing, which emerges in close proximity to the Fermi level. Additionally, several other Weyl points of both type-I and type-II appear at higher energies, as shown in Figure~\ref{fig:banddos}. However, these states lie far above the Fermi level and are therefore not discussed here, as they are unlikely to influence the transport properties significantly. To explore the topological characteristics of the type-I Weyl point near the Fermi level, we performed a detailed analysis combining Wannier-based tight-binding models and symmetry-based topological indicators. The initial step involves constructing a tight-binding Hamiltonian using maximally localized Wannier functions (MLWFs), derived from the first-principles calculations. The orbital analysis confirms that bands near the Fermi level are mainly derived from Nd $d$ orbitals.
Our self-consistent field (SCF) calculations show that the total magnetization of the system converges to zero, confirming a nonmagnetic ground state for Nd$_3$In. To ensure that this phase is indeed the most stable, we explicitly considered ferromagnetic (FM), antiferromagnetic (AFM), and nonmagnetic (NM) configurations. The total energies of these configurations are summarized in Table A1 in the Supplementary Information. The combination of zero magnetization and preserved time-reversal symmetry (TRS) motivates the calculation of $\mathbb{Z}_2$ topological invariants to determine the system’s topological classification \cite{fu2007topological}. These invariants are computed between bands 48 and 49 using the evolution of Wannier charge centers (WCCs) \cite{soluyanov2011computing,yu2011equivalent}, as implemented in the \textsc{WANNIERTOOLS} package. Figure~\ref{fig:WCC}(a)–(f) displays the evolution of the Wannier charge centers (WCCs) across the six time-reversal invariant momentum (TRIM) planes. As summarized in Table \ref{tab:z2index}, we observe an odd number of WCC crossings in the $k_x = 0.5$, $k_y = 0.0$, and $k_z = 0.0$ planes, which correspond to nontrivial $\mathbb{Z}_2$ indices. In contrast, the $k_x = 0.0$, $k_y = 0.5$, and $k_z = 0.5$ planes exhibit an even number of crossings, indicating trivial $\mathbb{Z}_2$ indices.
\begin{table}
	\centering
	\renewcommand{\arraystretch}{1.35}
	\large
	\begin{tabular}{lc}
		\toprule
		\textbf{Plane} & \textbf{Z\(_2\) index} \\
		\midrule
		k$_x = 0.0$, k$_y$--k$_z$ plane & 0 \\
		k$_x = 0.5$, k$_y$--k$_z$ plane & 1 \\
		k$_y = 0.0$, k$_x$--k$_z$ plane & 1 \\
		k$_y = 0.5$, k$_x$--k$_z$ plane & 0 \\
		k$_z = 0.0$, k$_x$--k$_y$ plane & 1 \\
		k$_z = 0.5$, k$_x$--k$_y$ plane & 0 \\
		\bottomrule
	\end{tabular}
	\vspace{0.8em}
	\caption{Z\(_2\) topological invariants for six time-reversal invariant momentum (TRIM) planes.}
	\label{tab:z2index}
\end{table}From this data, the strong and weak $\mathbb{Z}_2$ indices can be extracted, yielding $(\nu_0; \nu_1 \nu_2 \nu_3)=(1; 100)$,
which indicates that Nd$_3$In exhibits a strong topological phase. 
Additionally, employing VASP \cite{PhysRevB.54.11169} with \texttt{vasp2trace} and \texttt{CheckTopologicalMat} indicates that Nd$_3$In is an enforced semimetal with Fermi degeneracy (ESFD), supporting the presence of a nontrivial topological phase \cite{vergniory2019complete}.
To further elucidate the topological character, we compute the surface states on the (001) surface. These calculations reveal a rich spectrum of topological surface states (TSSs), as illustrated in Figures \ref{fig:ss}(a) and \ref{fig:ss}(c). The surface-projected band structure along the $\overline{\text{M}}$–$\overline{\Gamma}$–$\overline{\text{M}}$ direction, shown in Figure~\ref{fig:ss}(a), exhibits a prominent surface state crossing near the midpoint of $\overline{\Gamma}$–$\overline{\text{M}}$, within the energy window of 0.10–0.25 eV. This feature corresponds to the bulk type-I Weyl point observed along the $\Gamma$–M path. As discussed previously, the inclusion of SOC lifts some of the Weyl nodes and modifies the bulk bandstructure. Although partial gaps appear, no full insulating gap opens across the Brillouin zone. Consequently, Nd$_3$In retains its semimetallic character, and the corresponding surface states remain topologically nontrivial, as evidenced by a distinct linear crossing that connects the valence and conduction bands within the partially gapped region. This bridge is highlighted in the magnified inset of Figure~\ref{fig:ss}(a), underscoring the topological origin of the surface state. The existence of Weyl points in the bulk leads to topologically protected surface states called Fermi arcs that connect projections of Weyl points with opposite chirality \cite{nourizadeh2023emerging}. As shown in Figure \ref{fig:ss}(b), the Fermi arcs clearly connect the projected Weyl points, thereby confirming the topologically nontrivial nature of the material. We have also investigated the surface states along the $\overline{\text{M}}$–$\overline{\text{X}}$–$\overline{\text{M}}$ high-symmetry direction, as illustrated in Figure~\ref{fig:ss}(c). Notably, around the $\overline{\text{X}}$ point, Rashba splitting occurs near 0.30 eV, where the initially degenerate topological surface states (TSSs) split into two distinct branches \cite{zhang2025nodal}. A similar feature is evident along the $\overline{\Gamma}$–$\overline{\text{M}}$–$\overline{\Gamma}$ path near –0.75 eV, highlighting the presence of  Rashba-like TSSs that are of considerable interest in spintronics \cite{bihlmayer2022rashba,ishizaka2011giant}.\\
Before concluding, we briefly discuss the compositional variant In$_3$Nd, with the corresponding results summarized in Supplementary Materials \cite{SM}. The thermodynamic stability of both In$_3$Nd and Nd$_3$In was evaluated by constructing the convex hull of the Nd--In system as shown in Figure S5 of Supplementary Materials. In$_3$Nd crystallizes in the AuCu$_3$-type structure with a lattice constant of 4.721~\AA\ without SOC (4.684~\AA\ with SOC). We find that this material also exhibits semimetallic superconductivity. Orbital-projected band structures and projected density of states without and with SOC are shown in Figures S6 and S7, respectively. Figure S5 reveals several type-I Weyl points that are gapped out upon inclusion of SOC, resulting in a continuous gap between the highest valence band and the lowest conduction band. Notably, a Weyl point is observed precisely at the Fermi level along the $M$–$\Gamma$ path. The semimetallic nature of In$_3$Nd is further supported by the presence of Fermi arcs, as illustrated in Figure S8. From the superconducting perspective, In$_3$Nd exhibits a low transition temperature of $T_c = 0.72$~K, with a superconducting gap of 0.11~meV and an electron-phonon coupling constant $\lambda = 0.435$. This significant decrease in $T_c$ relative to Nd$_3$In is primarily attributed to a substantial decrease in the total electronic density of states at the Fermi level, which decreases from 6.19 to 1.51 states/eV due to the Nd–In site interchange. The mode-resolved electron-phonon coupling strength, Eliashberg spectral function, and electronic density of states for In$_3$Nd are detailed in Figure S8. Interestingly, this valency-exchanged phenomenon is also observed in the La--In system, where both La$_3$In and In$_3$La crystallize in the AuCu$_3$-type structure. While La$_3$In is a conventional superconductor \cite{sampathkumaran2014lattice}, its compositional counterpart In$_3$La exhibits semimetallic superconductivity \cite{wan2021structural,teicher20223d}.\\
The predictions made in this study offer clear pathways for experimental validation. ARPES offers direct momentum-resolved detection of topological surface states, while STM and quantum oscillations can spatially resolve these states and reveal signatures of the nontrivial Fermi surface topology \cite{yang2018quantum,soumyanarayanan2015momentum,shrestha2022nontrivial,lv2021experimental,ghosh2019observation}.
\section{Conclusions}
In summary, our comprehensive first-principles investigation identifies cubic Nd$_3$In as an unexplored semimetallic superconductor. Starting from its electronic structure, we find that the density of states (DOS) near the Fermi level is predominantly contributed by Nd $d$ orbitals. These orbitals form a three-band Fermi surface, playing a key role in the superconducting mechanism. The inclusion of spin–orbit coupling (SOC) reveals Kramers degeneracy and a gap opening. Phonon dispersion analysis reveals that low-frequency acoustic modes interact strongly with the electronic states, leading to a significant electron–phonon coupling (EPC) strength driven by pronounced Fermi nesting, consistent with strong-coupling superconductivity. To accurately capture its superconducting behavior, we solve the fully anisotropic Migdal–Eliashberg equations, yielding a superconducting transition temperature of 
$T_c \approx 14$ K at ambient pressure. Despite the pronounced anisotropy in the EPC, the superconducting gap remains single-band in nature. The computed gap to $T_c$ ratio, $2\Delta_0/k_BT_c$, significantly exceeds the BCS weak-coupling limit, further confirming the strong-coupling character of superconductivity in this material.
In addition to its superconducting properties, Nd$_3$In hosts nontrivial band topology, exhibiting characteristics of a Weyl semimetal. The topological analysis reveals surface Fermi arcs and a strong topological index with $\mathbb{Z}_2$ invariants $(\nu_0; \nu_1 \nu_2 \nu_3) = (1; 100)$, indicating a strong topological phase. The coexistence of strong-coupling superconductivity and nontrivial band topology makes Nd$_3$In a promising candidate for applications in quantum transport, topological quantum computing, and quantum information technology.

\bibliographystyle{apsrev4-2}
\bibliography{Final_manuscript}
\end{document}


\title{Supplementary Material for\\
		\vspace{1 cm} 
		Semimetallic Superconductivity in Cubic Nd$_3$In: A First-Principles  Insight into Indium-Based Compounds}
		\vspace{-1 cm}
	\author{Arafat Rahman}
	\affiliation{Department of Physics, University of Dhaka, Dhaka 1000, Bangladesh}
	\author{Alamgir Kabir}
	\affiliation{Department of Physics, University of Dhaka, Dhaka 1000, Bangladesh}
	\author{Tareq Mahmud}
	\affiliation{Department of Physics, University of Dhaka, Dhaka 1000, Bangladesh}
	
	\maketitle
	\begin{figure}[H]
		\centering
		\includegraphics[width=0.6\linewidth]{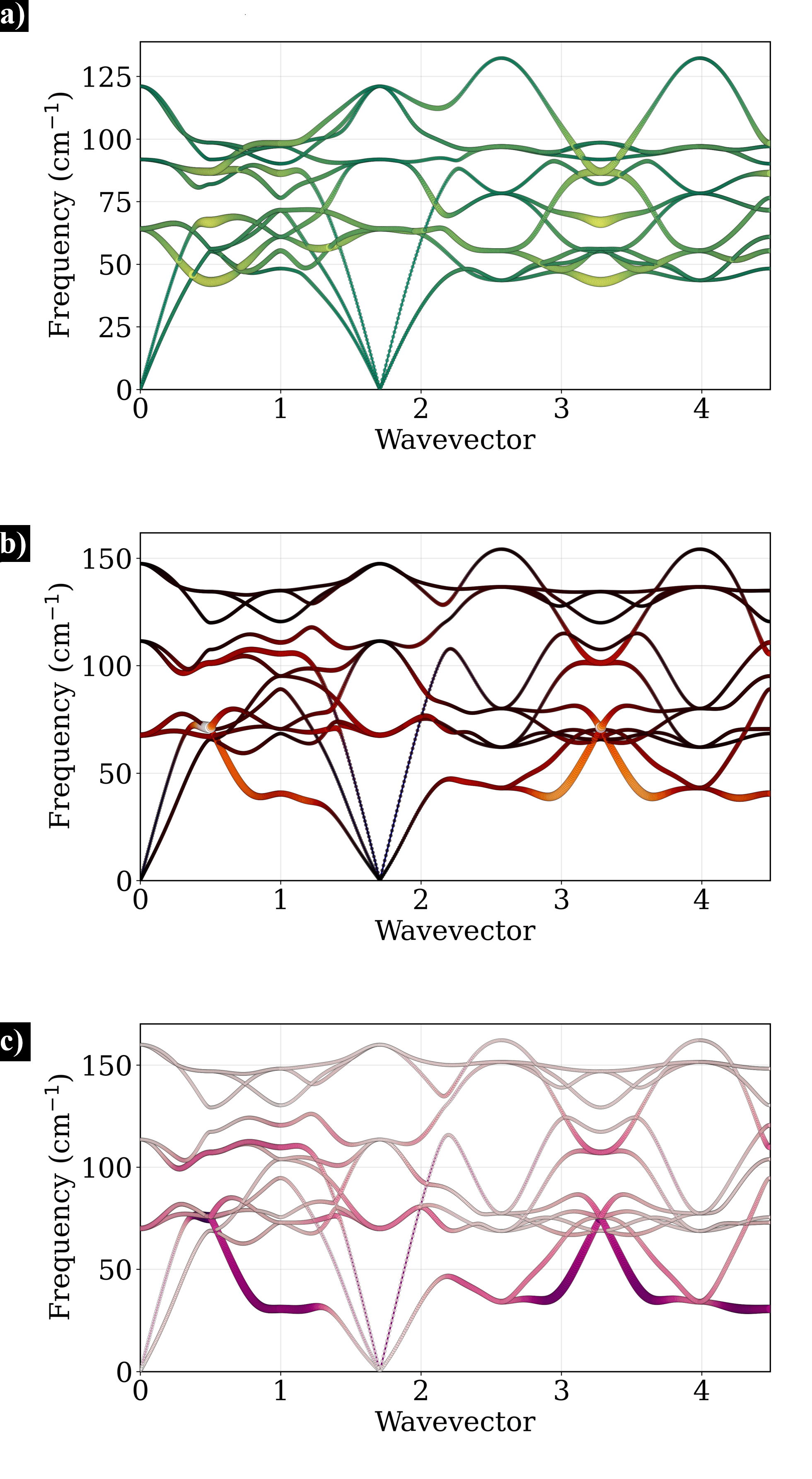}
		\caption{Phonon dispersion relations of Nd$_3$In under different hydrostatic pressures. Fat bands indicate modes with strong electron–phonon coupling. (a) At ambient pressure (0 GPa), the phonon spectrum shows distinct acoustic and optical branches. (b) At 10 GPa, acoustic phonon branches soften, reflecting pressure-induced lattice changes. (c) At 15 GPa, further softening is observed in the low-frequency region. The electron-phonon coupling constant $\lambda$ increases with pressure, suggesting enhanced superconducting interactions.}
		\label{fig:phonall}
	\end{figure}
	
	\begin{figure}[H]
		\centering
		\includegraphics[width=\linewidth]{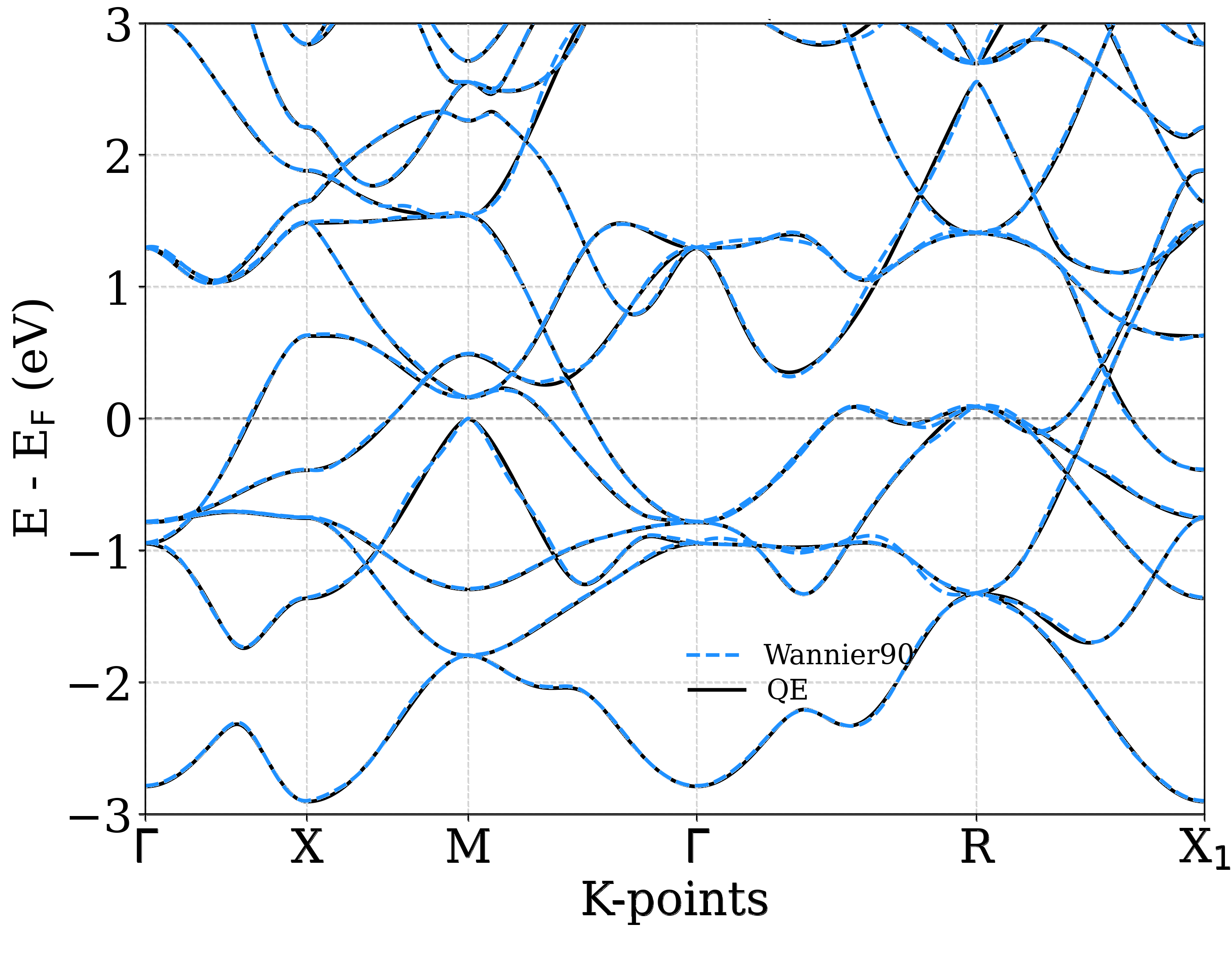}
		\caption{Electronic band structure of Nd$_3$In without spin–orbit coupling. The Wannier90-interpolated bands are shown as blue dashed lines; DFT bands as solid black lines.}
		\label{fig:bandcomp}
	\end{figure}
	
	\begin{figure}[H]
		\centering
		\includegraphics[width=0.7\linewidth]{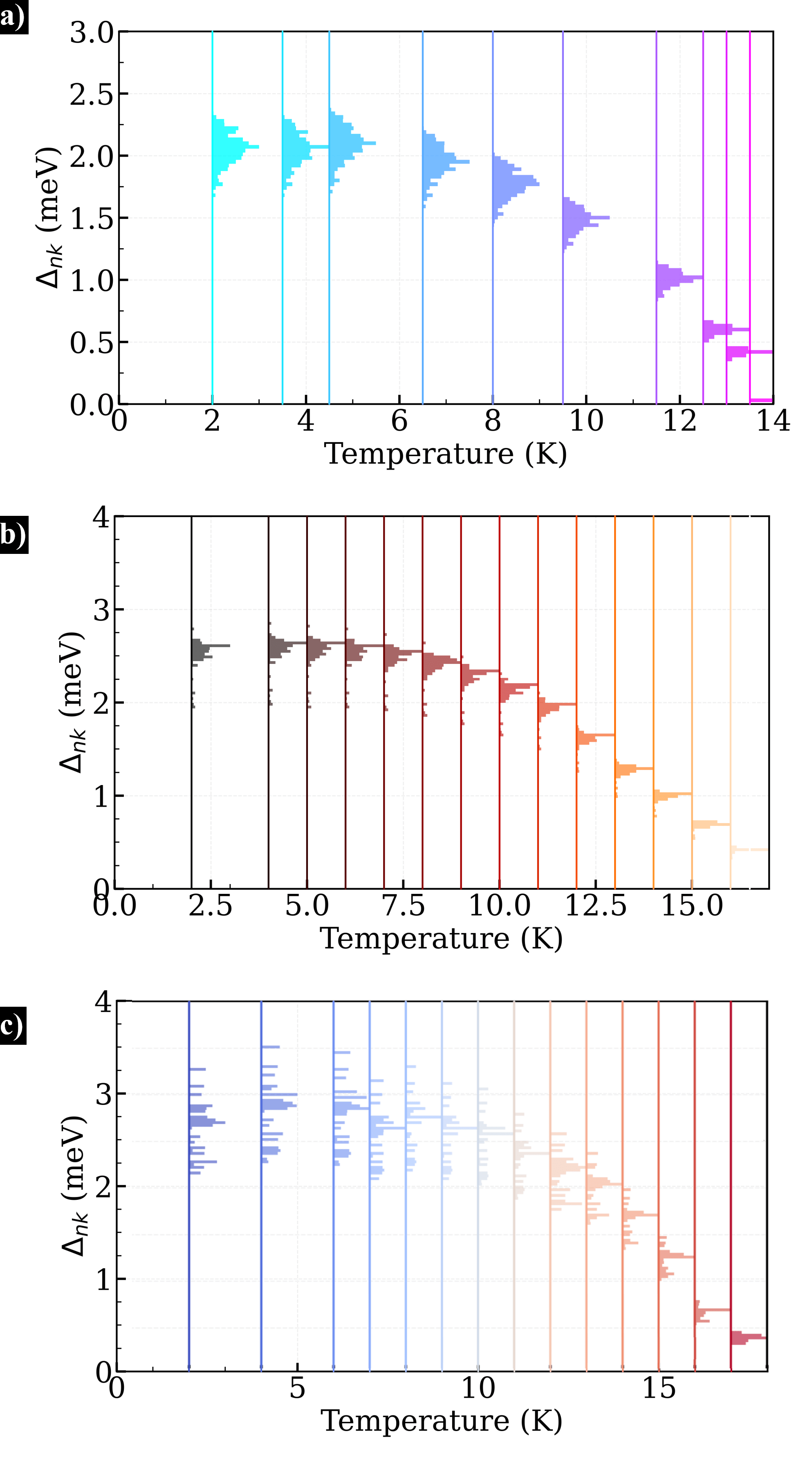}
		\caption{Temperature dependence of the anisotropic superconducting gap on the Fermi surface of Nd$_3$In. (a) At 0 GPa, the gap closes near 14 K. (b) At 10 GPa, it persists up to ~16.5 K. (c) At 15 GPa, it remains open up to ~18 K, confirming pressure-induced enhancement of $T_c$.}
		\label{fig:anisoall}
	\end{figure}
	
	\begin{figure}[H]
		\centering
		\includegraphics[width=0.5\linewidth]{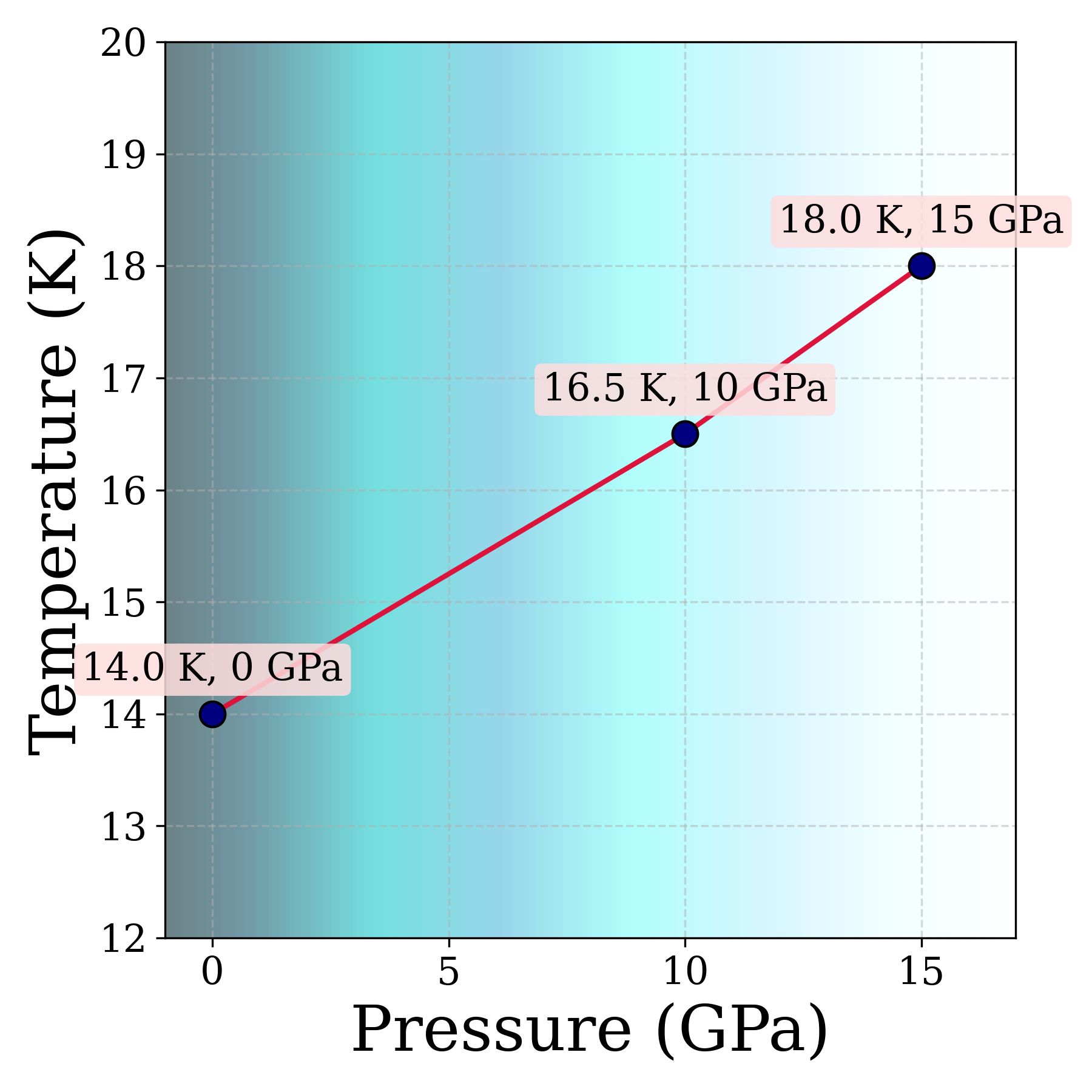}
		\caption{Superconducting transition temperature $T_c$ of Nd$_3$In at different pressures (0, 10, and 15 GPa) from anisotropic Eliashberg theory. A clear enhancement of $T_c$ with increasing pressure is observed.}
		\label{fig:TP}
	\end{figure}
	\begin{table}[H]
	\renewcommand{\thetable}{A1}
	\centering
	\caption{Total energies of Nd$_3$In for different magnetic configurations.}
	\label{tab:S1}
	\resizebox{\textwidth}{!}{%
		\begin{tabular}{|l|c|c|c|c|}
			\hline
			\textbf{Magnetic Config.} & \textbf{Total Energy (Ry)} & \textbf{Total Magnetization ($\mu_B$} & $\Delta E$ w.r.t NM (Ry) & $\Delta E$ w.r.t NM (eV) \\
			\hline
			Nonmagnetic (NM) & -521.37877592 & 0.00 & 0.00000 & 0.0000 \\
			\hline
			Antiferromagnetic & -521.37823121 & 0.00 & +0.00054471 & +0.0074 \\
			\hline
			Ferromagnetic (FM) & -521.00284678 & 9.00 & +0.37592914 & +5.11 \\
			\hline
		\end{tabular}%
	}
	\end{table}
	
	\begin{figure}[H]
		\centering
		\includegraphics[width=0.5\linewidth]{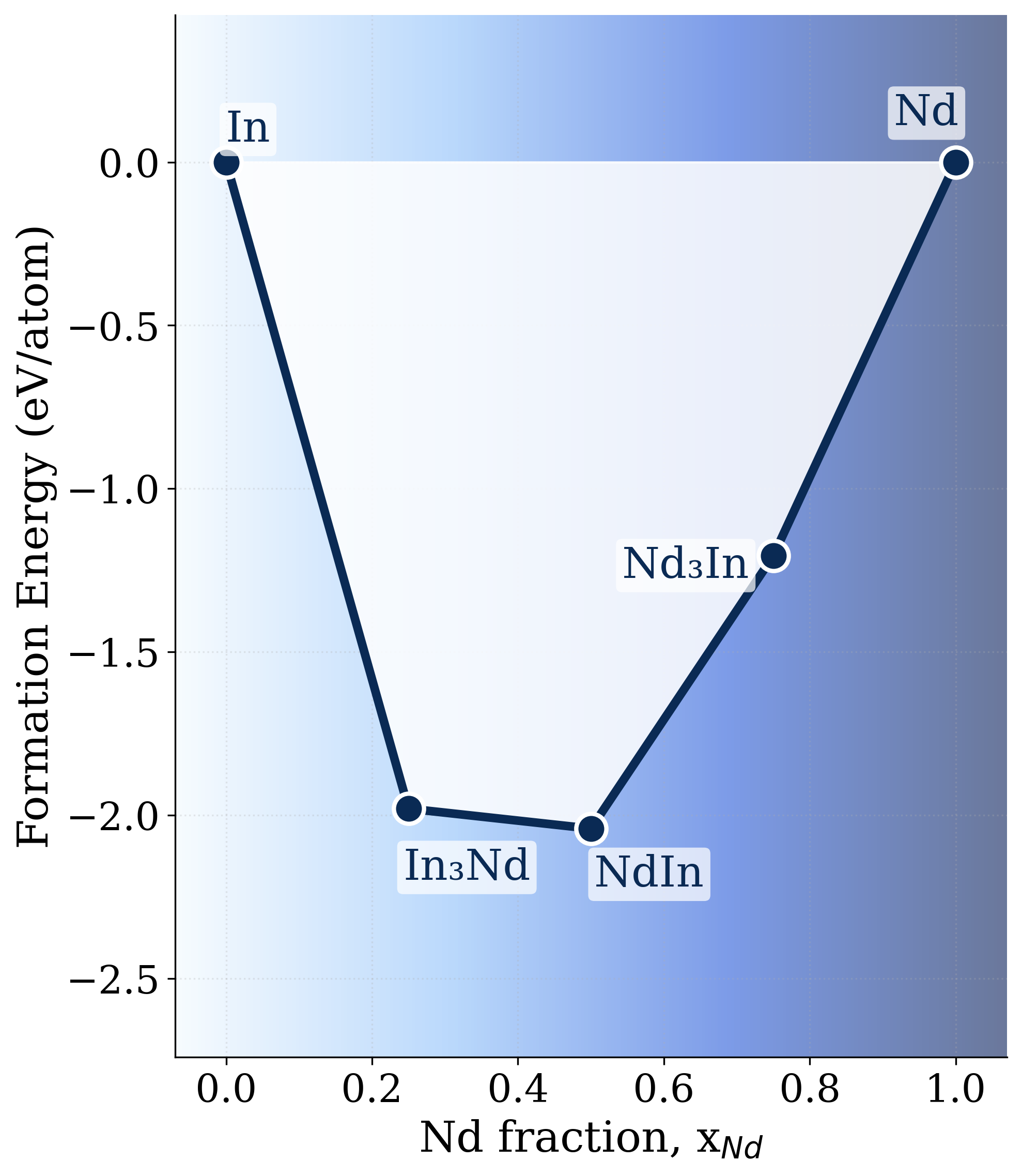}
		\caption{Convex hull of the Nd–In binary system showing the stability of In$_3$Nd and Nd$_3$In.}
		\label{fig:convex}
	\end{figure}
	
	\begin{figure}[H]
		\centering
		\includegraphics[width=0.9\linewidth]{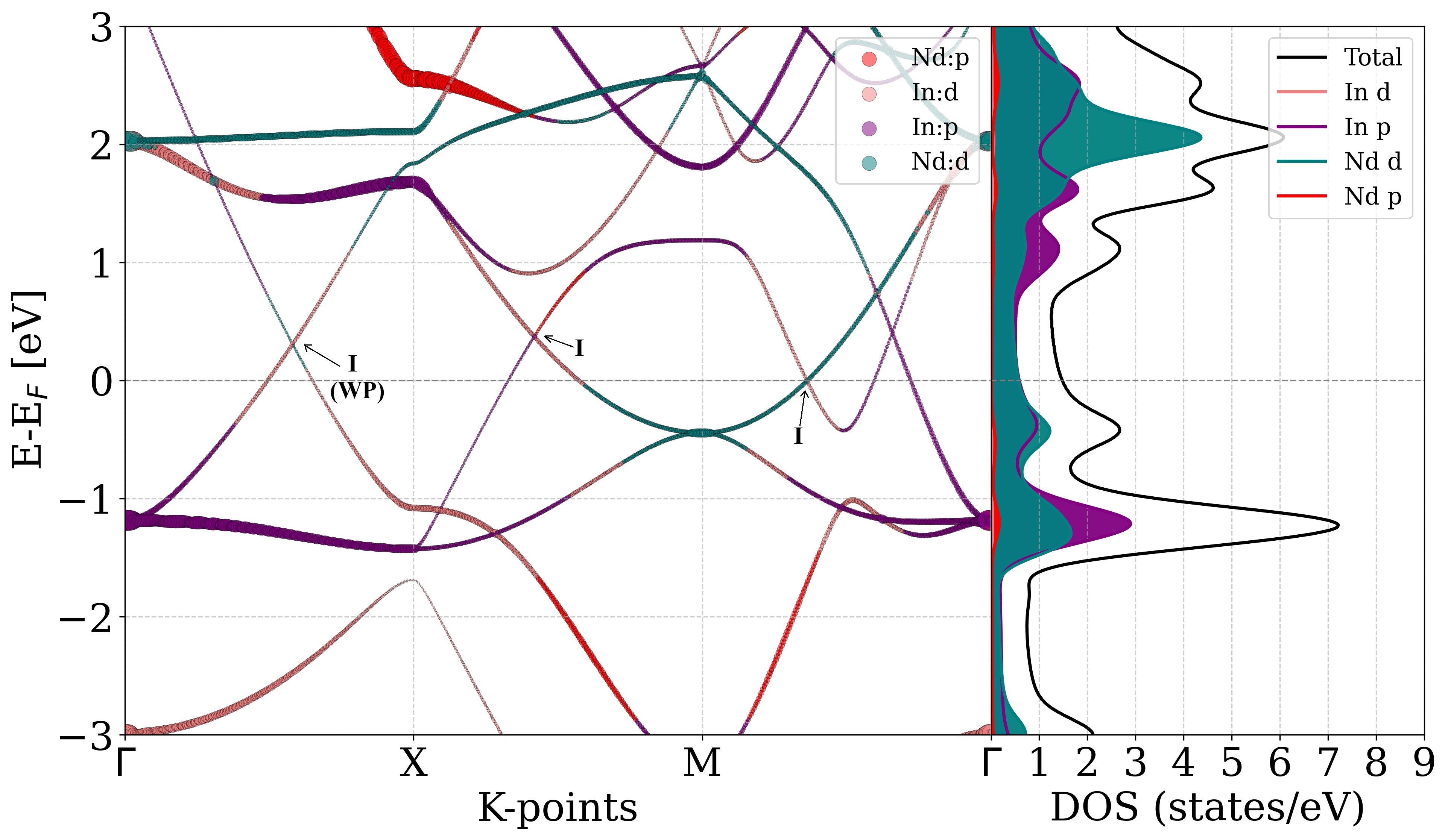}
		\caption{Orbital-projected band structure and density of states (PDOS) for In$_3$Nd without spin–orbit coupling. Scatter size indicates orbital contribution. The total DOS at the Fermi level is 1.51 states/eV, indicating limited electronic states.}
		\label{fig:NdIn3_band}
	\end{figure}
	
	\begin{figure}[H]
		\centering
		\includegraphics[width=\linewidth]{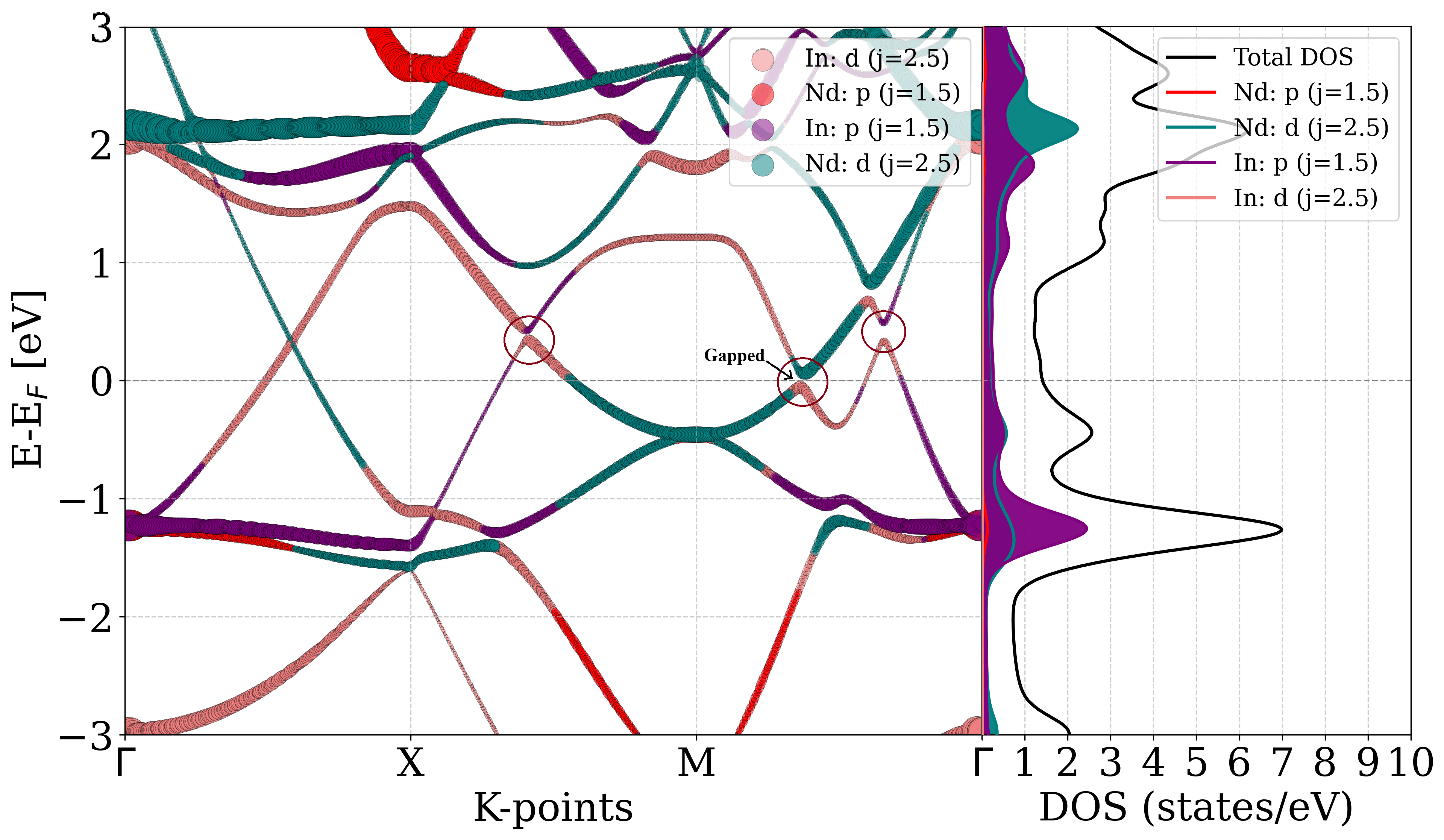}
		\caption{Band structure and PDOS of In$_3$Nd with spin–orbit coupling. Orbital projections shown by scatter size. DOS at the Fermi level slightly decreases to ~1.38 states/eV. Several band gaps appear due to SOC.}
		\label{fig:NdIn3_relband}
	\end{figure}
	
	\begin{figure}[H]
		\centering
		\includegraphics[width=0.5\linewidth]{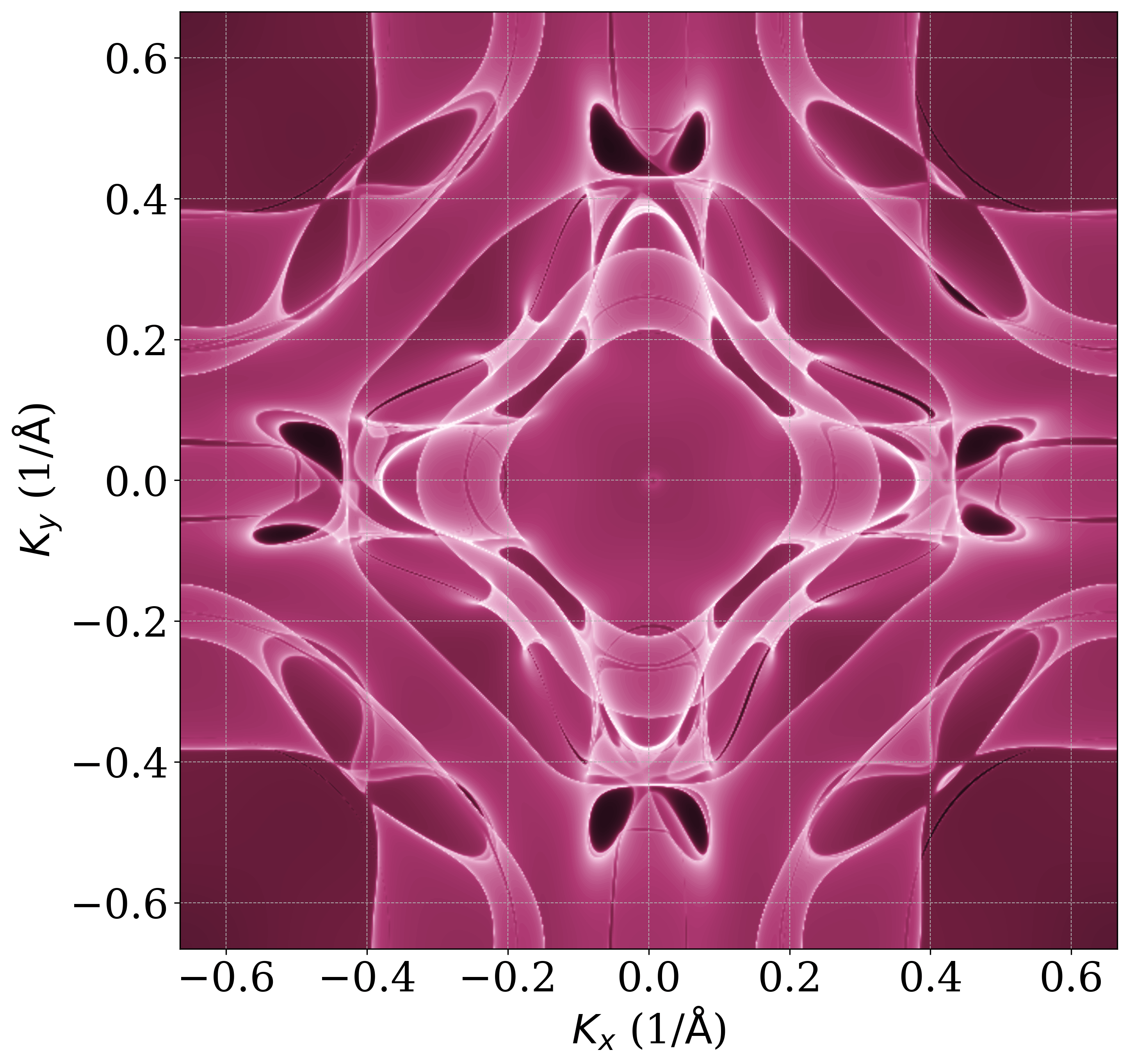}
		\caption{Fermi arcs in In$_3$Nd, indicating semimetallic character.}
		\label{fig:NdIn3_arc}
	\end{figure}
	
	\begin{figure}[H]
		\centering
		\includegraphics[width=\linewidth]{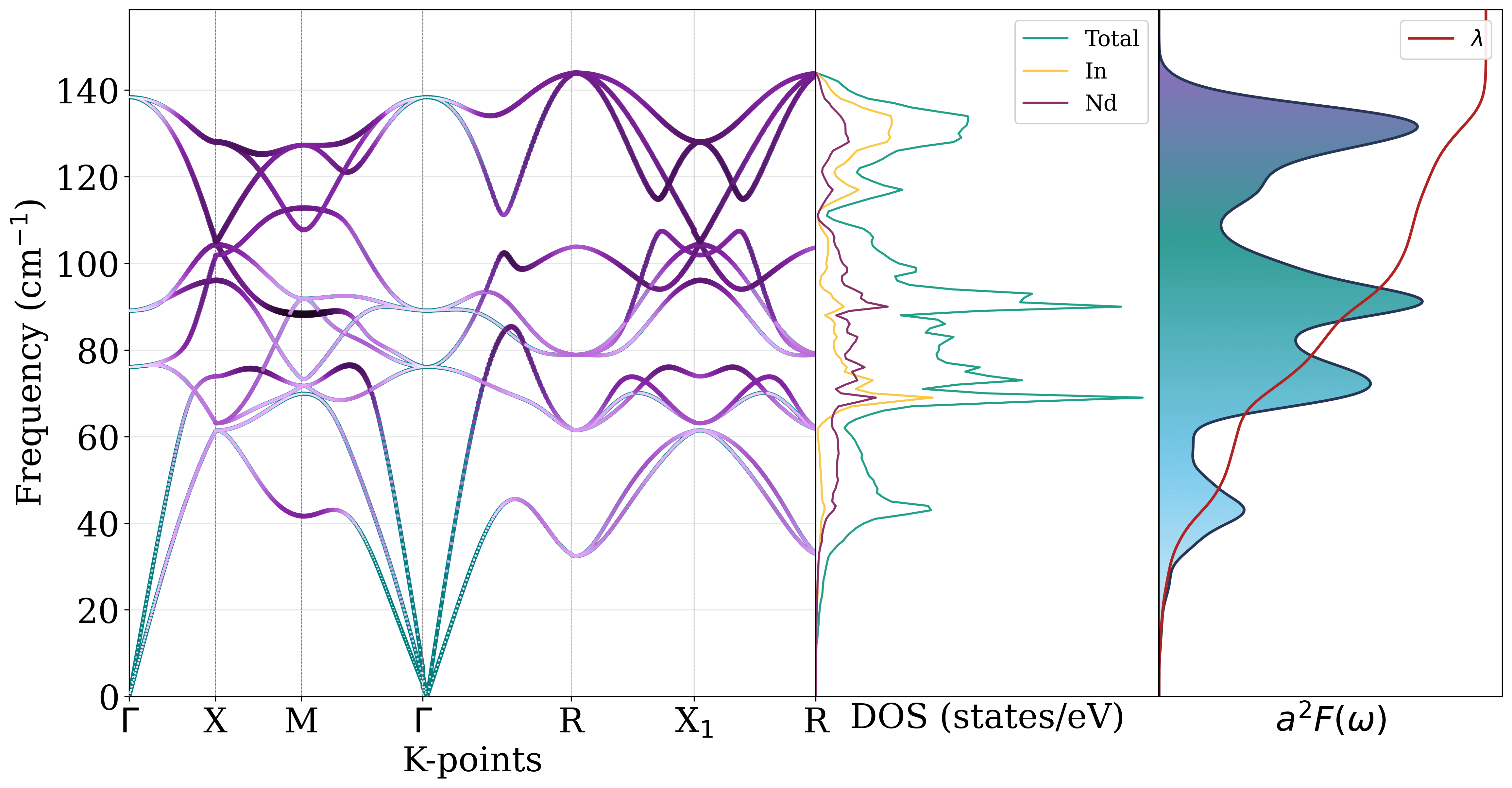}
		\caption{Phononic and superconducting properties of In$_3$Nd. Left: phonon dispersion weighted by mode-resolved EPC $\lambda_{\mathbf{q}\nu}$. Middle: atom-projected phonon DOS. Right: Eliashberg spectral function $\alpha^2F(\omega)$ and integrated EPC $\lambda(\omega)$.}
		\label{fig:NdIn3_epw}
	\end{figure}